\pdfoutput=1
\documentclass[sigconf,nonacm,balance=false]{acmart}

\usepackage{flushend}
\usepackage{tikz}
\usetikzlibrary{arrows.meta,calc}
\usepackage{pgfplots}
\pgfplotsset{compat=1.18}
\usepgfplotslibrary{groupplots}

\hypersetup{%
  pdfauthor={Zhi Kou, Zhangming Chan, Yichao Liu, Jingchen Gao, Chenxi Wang, Jianqi Xie, Yan Jiang, Zhenyuan Lai, Chaochao Zhao, Wenchao Wang, Dagui Chen, Jian Xu, Bo Zheng},
  pdftitle={OmniPreRank: Progressive Computation Reuse for Pre-Ranking as a Step toward Unified Ranking},
  pdfkeywords={Pre-ranking, Unified Ranking, Cascade Ranking, Computation Reuse, Online Advertising}%
}

\AtBeginDocument{%
  \setlength{\abovedisplayskip}{6pt plus 1.5pt minus 1.5pt}%
  \setlength{\belowdisplayskip}{6pt plus 1.5pt minus 1.5pt}%
  \setlength{\abovedisplayshortskip}{3pt plus 1pt minus 1pt}%
  \setlength{\belowdisplayshortskip}{3pt plus 1pt minus 1pt}%
  \setlength{\textfloatsep}{10pt plus 2pt minus 2pt}%
  \setlength{\dbltextfloatsep}{12pt plus 2pt minus 2pt}%
  \setlength{\floatsep}{8pt plus 2pt minus 2pt}%
}

\begin{document}
\title[OmniPreRank]{OmniPreRank: Progressive Computation Reuse for Pre-Ranking as a Step toward Unified Ranking}

\author{Zhi Kou}
\authornote{These authors contributed equally.}
\email{kouzhi.kz@alibaba-inc.com}
\author{Zhangming Chan}
\authornotemark[1]
\email{zhangming.czm@alibaba-inc.com}
\author{Yichao Liu}
\author{Jingchen Gao}
\author{Chenxi Wang}
\author{Jianqi Xie}
\author{Yan Jiang}
\author{Zhenyuan Lai}
\author{Chaochao Zhao}
\author{Wenchao Wang}
\author{Dagui Chen}
\author{Jian Xu}
\author{Bo Zheng}
\authornote{Corresponding author.}
\email{bozheng@alibaba-inc.com}
\affiliation{%
  \institution{Taobao \& Tmall Group of Alibaba}
  \city{Beijing}
  \country{China}}

\renewcommand{\shortauthors}{Kou et al.}

\makeatletter
\if@ACM@anonymous\else
  \newcount\omni@authorcount
  \def\@mkauthors{%
    \begingroup
    \gdef\@currentauthors{}%
    \omni@authorcount=0\relax
    \def\@author##1{%
      \advance\omni@authorcount by 1\relax
      \ifnum\omni@authorcount=1\relax
        \gdef\@currentauthors{##1}%
      \else\ifnum\omni@authorcount=8\relax
        \g@addto@macro\@currentauthors{,\\##1}%
      \else
        \g@addto@macro\@currentauthors{, ##1}%
      \fi\fi}%
    \def\email##1##2{}%
    \def\affiliation##1##2{}%
    \def\and{}%
    \addresses
    \global\setbox\mktitle@bx=\vbox{%
      \unvbox\mktitle@bx
      \centering\medskip
      {\@authorfont\@currentauthors\par}
      \smallskip
      {\@affiliationfont Taobao \& Tmall Group of Alibaba, Beijing, China\par}
      {\@affiliationfont
        \{\href{mailto:kouzhi.kz@alibaba-inc.com}{kouzhi.kz},
        \href{mailto:zhangming.czm@alibaba-inc.com}{zhangming.czm},
        \href{mailto:bozheng@alibaba-inc.com}{bozheng}\}@alibaba-inc.com\par}
      \bigskip}%
    \endgroup}
\fi
\makeatother

\begin{abstract}
Unified ranking, which replaces the retrieval-and-ranking cascade with
one model, is gaining attention in recommendation, but industrial
advertising systems tightly couple each stage with bidding, candidate
filtering, and business rules, making wholesale replacement costly. We
pursue an incremental path: stages retain their decision roles while
their models and computation are progressively merged, so later stages
continue earlier computation rather than restarting. We instantiate
this design as \emph{OmniPreRank}, a multi-task Transformer that merges
computation across a filter within pre-ranking. Shallow and Deep heads
at two depths of a shared backbone each predict a ranking signal and
the transaction value used for bidding. The shallow round scores all candidates; after bidding and engine-side filtering, the deep round
resumes from retained candidates' intermediate states. GPU state
caching and request routing enable continuation across requests, while
low-latency data exchange between the prediction and GPU bidding services
keeps the second bidding call within the latency budget. Deployed with a
$16{,}000\to2{,}000\to400$ funnel, OmniPreRank improves consistency
with fine-ranking at both depths and increases RPM by
2.14\%, retaining 91.1\% of the RPM gain of full-candidate deep scoring,
with 36.7\% fewer prediction machines than the complete production fleet
serving separate LTR and GMV models under equal request load.
Fully deployed in production, OmniPreRank represents a practical step
toward unified ranking, enabling successive decisions to share
one model and progressively reuse computation.
\end{abstract}


\maketitle

\section{Introduction}
\label{sec:introduction}

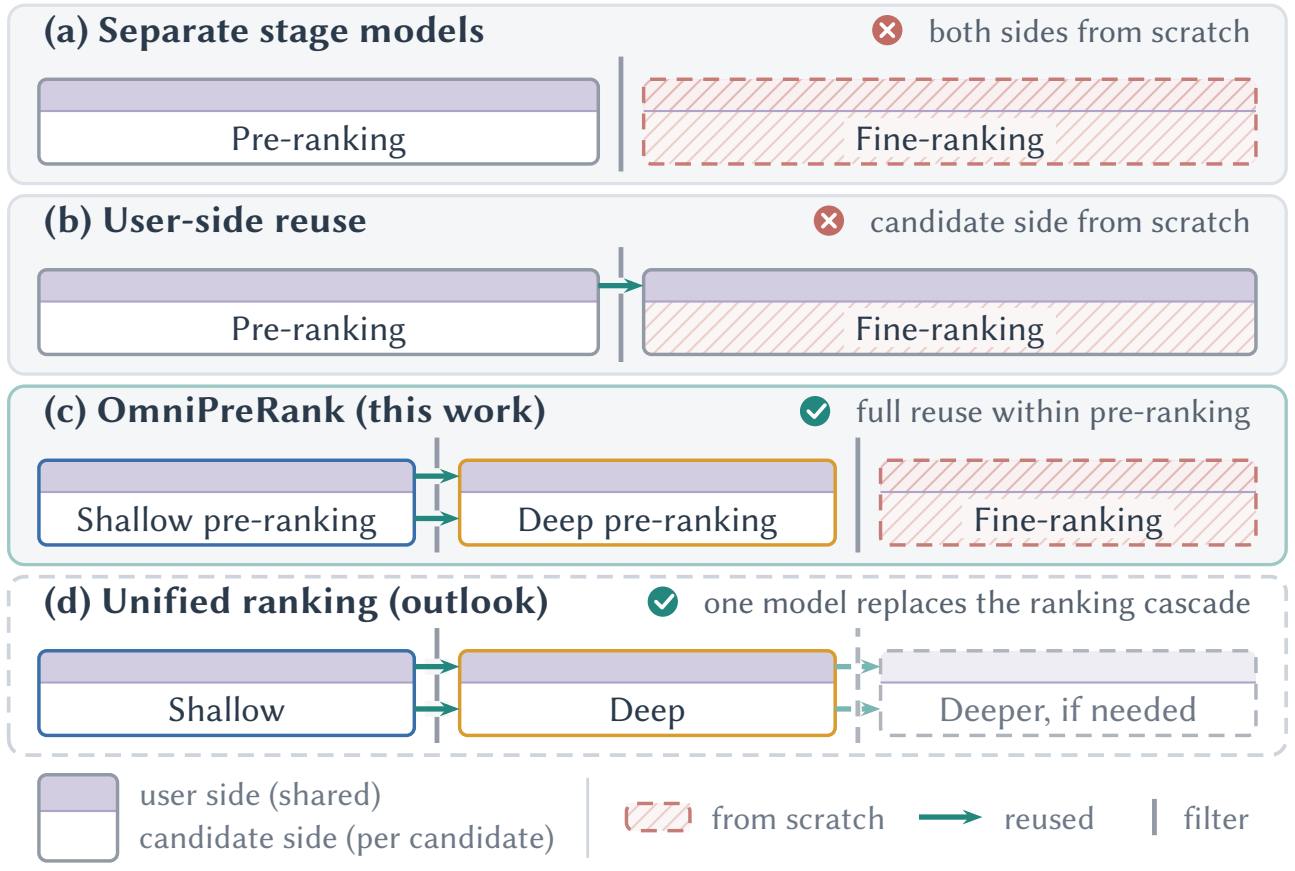
\begin{figure}[t]
\centering
\begingroup
\definecolor{omniInk}{HTML}{2B3A4A}
\definecolor{omniMuted}{HTML}{6E7987}
\definecolor{omniLine}{HTML}{CDD4DC}
\definecolor{omniShallow}{HTML}{3A6EA8}
\definecolor{omniDeep}{HTML}{D99A2E}
\definecolor{omniState}{HTML}{23877E}
\definecolor{omniBehavior}{HTML}{8C7BB0}
\definecolor{omniRecompute}{HTML}{B5524A}
\begin{tikzpicture}[x=1cm,y=1cm,
  font=\sffamily\fontsize{6.8}{8.2}\selectfont,text=omniInk,
  >={Stealth[length=1.3mm,width=.95mm]},
  lbl/.style={text height=1.55ex,text depth=.35ex,inner sep=0pt},
  title/.style={font=\sffamily\bfseries\fontsize{7.2}{8.6}\selectfont,anchor=base west},
  note/.style={font=\sffamily\fontsize{6.3}{7.4}\selectfont,text=omniMuted,align=center},
  card/.style={fill=omniLine!20,rounded corners=3pt},
  cand/.style={rounded corners=1.5pt,line width=.65pt,fill=white},
  sepl/.style={draw=omniMuted!75,line width=.9pt},
  keep/.style={->,draw=omniState,line width=1pt},
  usr/.style={fill=omniBehavior!38,draw=omniBehavior!85,line width=.5pt,rounded corners=.8pt},
  redo/.style={draw=omniRecompute!75,densely dashed,line width=.65pt,rounded corners=1.5pt}]
\path[use as bounding box] (0,0) rectangle (8.45,5.9);
\def\h{.2}
\newcommand{\hatchbox}[4]{
  \fill[omniRecompute!5,rounded corners=1.5pt] (#1,#2) rectangle (#3,#4);
  \begin{scope}
    \clip[rounded corners=1.5pt] (#1,#2) rectangle (#3,#4);
    \foreach \k in {0,...,26} {
      \draw[omniRecompute!28,line width=.4pt] ({#1-.6+.11*\k},#2) -- ++(.6,.6);}
  \end{scope}
  \draw[redo] (#1,#2) rectangle (#3,#4);}
\newcommand{\hatchlines}[5]{
  \begin{scope}\clip (#1,#2) rectangle (#3,#4);
    \foreach \k in {0,...,46} {\draw[#5,line width=.4pt] ({#1-.6+.11*\k},#2) -- ++(.6,.6);}
  \end{scope}}
\newcommand{\rowcard}[2]{\filldraw[card,draw=omniLine!70,line width=.6pt] (0,#1-.38) rectangle (8.45,#1+.8);
  \node[title] at (.12,#1+.56) {#2};}
\newcommand{\filt}[2]{\draw[sepl] (#2,#1-.3) -- (#2,#1+.46);}
\newcommand{\carry}[4]{
  \draw[draw=omniLine!20,line width=3pt] (#2-.08,#1) -- (#2+.08,#1);
  \draw[keep] (#3,#1) -- (#4,#1);}
\newcommand{\verdict}[3]{
  \node[note,text=omniInk!80,anchor=base east] (v) at (8.33,#1) {#3};
  \fill[#2] ([xshift=-1.6mm,yshift=.75mm]v.base west) circle (.1);}
\newcommand{\bad}[2]{\verdict{#1}{omniRecompute!85}{#2}
  \draw[white,line width=.7pt,line cap=round] ([xshift=-2mm,yshift=.35mm]v.base west) -- ++(.08,.08) ([xshift=-2mm,yshift=1.15mm]v.base west) -- ++(.08,-.08);}
\newcommand{\good}[2]{\verdict{#1}{omniState}{#2}
  \draw[white,line width=.7pt,line cap=round,line join=round] ([xshift=-2.05mm,yshift=.75mm]v.base west) -- ++(.04,-.04) -- ++(.07,.08);}
\rowcard{4.92}{(a) Separate stage models}
\bad{5.48}{both sides from scratch}
\begin{scope}\clip[rounded corners=1.5pt] (0.2,4.67) rectangle (3.9,5.23);
\fill[white] (0.2,4.67) rectangle (3.9,5.02);
\fill[omniBehavior!38] (0.2,5.02) rectangle (3.9,5.23);
\end{scope}
\draw[draw=omniBehavior!70,line width=.4pt] (0.2,5.02) -- (3.9,5.02);
\draw[cand,fill=none,draw=omniMuted!75] (0.2,4.67) rectangle (3.9,5.23);
\node[lbl] at (2.05,4.845) {Pre-ranking};
\begin{scope}\clip[rounded corners=1.5pt] (4.2,4.67) rectangle (8.25,5.23);
\fill[white] (4.2,4.67) rectangle (8.25,5.02);
\fill[omniBehavior!38] (4.2,5.02) rectangle (8.25,5.23);
\fill[omniRecompute!5] (4.2,4.67) rectangle (8.25,5.02);
\hatchlines{4.2}{4.67}{8.25}{5.02}{omniRecompute!28}
\fill[omniRecompute!8] (4.2,5.02) rectangle (8.25,5.23);
\hatchlines{4.2}{5.02}{8.25}{5.23}{omniRecompute!40}
\end{scope}
\draw[draw=omniBehavior!70,line width=.4pt] (4.2,5.02) -- (8.25,5.02);
\draw[redo] (4.2,4.67) rectangle (8.25,5.23);
\node[lbl,fill=omniRecompute!5,inner xsep=2pt] at (6.225,4.845) {Fine-ranking};
\filt{4.92}{4.05}

\rowcard{3.66}{(b) User-side reuse}
\bad{4.22}{candidate side from scratch}
\begin{scope}\clip[rounded corners=1.5pt] (0.2,3.41) rectangle (3.9,3.97);
\fill[white] (0.2,3.41) rectangle (3.9,3.76);
\fill[omniBehavior!38] (0.2,3.76) rectangle (3.9,3.97);
\end{scope}
\draw[draw=omniBehavior!70,line width=.4pt] (0.2,3.76) -- (3.9,3.76);
\draw[cand,fill=none,draw=omniMuted!75] (0.2,3.41) rectangle (3.9,3.97);
\node[lbl] at (2.05,3.585) {Pre-ranking};
\begin{scope}\clip[rounded corners=1.5pt] (4.2,3.41) rectangle (8.25,3.97);
\fill[white] (4.2,3.41) rectangle (8.25,3.76);
\fill[omniBehavior!38] (4.2,3.76) rectangle (8.25,3.97);
\fill[omniRecompute!5] (4.2,3.41) rectangle (8.25,3.76);
\hatchlines{4.2}{3.41}{8.25}{3.76}{omniRecompute!28}
\end{scope}
\draw[draw=omniBehavior!70,line width=.4pt] (4.2,3.76) -- (8.25,3.76);
\draw[cand,fill=none,draw=omniMuted!75] (4.2,3.41) rectangle (8.25,3.97);
\node[lbl,fill=omniRecompute!5,inner xsep=2pt] at (6.225,3.585) {Fine-ranking};
\filt{3.66}{4.05}
\carry{3.865}{4.05}{3.9}{4.2}

\rowcard{2.4}{(c) OmniPreRank (this work)}
\draw[omniState!45,line width=.6pt,rounded corners=3pt] (0,2.02) rectangle (8.45,3.2);
\good{2.96}{full reuse within pre-ranking}
\begin{scope}\clip[rounded corners=1.5pt] (0.2,2.15) rectangle (2.683,2.71);
\fill[white] (0.2,2.15) rectangle (2.683,2.5);
\fill[omniBehavior!38] (0.2,2.5) rectangle (2.683,2.71);
\end{scope}
\draw[draw=omniBehavior!70,line width=.4pt] (0.2,2.5) -- (2.683,2.5);
\draw[cand,fill=none,draw=omniShallow] (0.2,2.15) rectangle (2.683,2.71);
\node[lbl] at (1.442,2.325) {Shallow pre-ranking};
\begin{scope}\clip[rounded corners=1.5pt] (2.983,2.15) rectangle (5.467,2.71);
\fill[white] (2.983,2.15) rectangle (5.467,2.5);
\fill[omniBehavior!38] (2.983,2.5) rectangle (5.467,2.71);
\end{scope}
\draw[draw=omniBehavior!70,line width=.4pt] (2.983,2.5) -- (5.467,2.5);
\draw[cand,fill=none,draw=omniDeep] (2.983,2.15) rectangle (5.467,2.71);
\node[lbl] at (4.225,2.325) {Deep pre-ranking};
\begin{scope}\clip[rounded corners=1.5pt] (5.767,2.15) rectangle (8.25,2.71);
\fill[white] (5.767,2.15) rectangle (8.25,2.5);
\fill[omniBehavior!38] (5.767,2.5) rectangle (8.25,2.71);
\fill[omniRecompute!5] (5.767,2.15) rectangle (8.25,2.5);
\hatchlines{5.767}{2.15}{8.25}{2.5}{omniRecompute!28}
\fill[omniRecompute!8] (5.767,2.5) rectangle (8.25,2.71);
\hatchlines{5.767}{2.5}{8.25}{2.71}{omniRecompute!40}
\end{scope}
\draw[draw=omniBehavior!70,line width=.4pt] (5.767,2.5) -- (8.25,2.5);
\draw[redo] (5.767,2.15) rectangle (8.25,2.71);
\node[lbl,fill=omniRecompute!5,inner xsep=2pt] at (7.008,2.325) {Fine-ranking};
\filt{2.4}{2.833}
\filt{2.4}{5.617}
\carry{2.605}{2.833}{2.683}{2.983}
\carry{2.325}{2.833}{2.683}{2.983}

\draw[omniLine,line width=.7pt,densely dashed,rounded corners=3pt] (0,0.76) rectangle (8.45,1.94);
\node[title] at (.12,1.7) {(d) Unified ranking (outlook)};
\good{1.7}{one model replaces the ranking cascade}
\begin{scope}\clip[rounded corners=1.5pt] (0.2,0.89) rectangle (2.683,1.45);
\fill[white] (0.2,0.89) rectangle (2.683,1.24);
\fill[omniBehavior!38] (0.2,1.24) rectangle (2.683,1.45);
\end{scope}
\draw[draw=omniBehavior!70,line width=.4pt] (0.2,1.24) -- (2.683,1.24);
\draw[cand,fill=none,draw=omniShallow] (0.2,0.89) rectangle (2.683,1.45);
\node[lbl] at (1.442,1.065) {Shallow};
\begin{scope}\clip[rounded corners=1.5pt] (2.983,0.89) rectangle (5.467,1.45);
\fill[white] (2.983,0.89) rectangle (5.467,1.24);
\fill[omniBehavior!38] (2.983,1.24) rectangle (5.467,1.45);
\end{scope}
\draw[draw=omniBehavior!70,line width=.4pt] (2.983,1.24) -- (5.467,1.24);
\draw[cand,fill=none,draw=omniDeep] (2.983,0.89) rectangle (5.467,1.45);
\node[lbl] at (4.225,1.065) {Deep};
\begin{scope}\clip[rounded corners=1.5pt] (5.767,0.89) rectangle (8.25,1.45);
\fill[white] (5.767,0.89) rectangle (8.25,1.24);
\fill[omniBehavior!14] (5.767,1.24) rectangle (8.25,1.45);
\end{scope}
\draw[draw=omniBehavior!70,line width=.4pt] (5.767,1.24) -- (8.25,1.24);
\draw[cand,fill=none,draw=omniMuted!55,densely dashed] (5.767,0.89) rectangle (8.25,1.45);
\node[lbl,text=omniMuted] at (7.008,1.065) {Deeper, if needed};
\filt{1.14}{2.833}
\draw[omniMuted!55,line width=.9pt,densely dashed] (5.617,0.84) -- (5.617,1.6);
\carry{1.345}{2.833}{2.683}{2.983}
\carry{1.065}{2.833}{2.683}{2.983}
\draw[->,draw=omniState!60,line width=1pt,densely dashed] (5.467,1.345) -- (5.767,1.345);
\draw[->,draw=omniState!60,line width=1pt,densely dashed] (5.467,1.065) -- (5.767,1.065);

\begin{scope}[shift={(0,0.36)},every node/.append style={note,text height=1.6ex,text depth=.4ex,inner sep=1pt}]
  \begin{scope}\clip[rounded corners=1.5pt] (.2,-.3) rectangle (.72,.27);
    \fill[white] (.2,-.3) rectangle (.72,.03); \fill[omniBehavior!38] (.2,.03) rectangle (.72,.27);
  \end{scope}
  \draw[omniBehavior!70,line width=.4pt] (.2,.03) -- (.72,.03);
  \draw[cand,fill=none,draw=omniMuted!75] (.2,-.3) rectangle (.72,.27);
  \node[anchor=west] at (.82,.15) {user side (shared)};
  \node[anchor=west] at (.82,-.135) {candidate side (per candidate)};
  \draw[omniLine,line width=.5pt] (3.83,-.28) -- (3.83,.25);
  \node[anchor=east] (t3) at (8.25,-.01) {filter};
  \draw[sepl] ([xshift=-1.5mm]t3.west |- 0,-.13) -- ++(0,.24);
  \node[anchor=east] (t2) at ([xshift=-5mm]t3.west) {reused};
  \draw[keep] ([xshift=-5.2mm]t2.west |- 0,-.01) -- ++(.42,0);
  \node[anchor=east] (t1) at ([xshift=-7mm]t2.west) {from scratch};
  \path ([xshift=-5.2mm]t1.west |- 0,-.1) coordinate (hr);
  \fill[omniRecompute!5] (hr) rectangle ++(.42,.18);
  \begin{scope}\clip (hr) rectangle ++(.42,.18);
    \foreach \k in {0,...,6} {\draw[omniRecompute!28,line width=.4pt] ($(hr)+(-.2+.11*\k,0)$) -- ++(.2,.2);}
  \end{scope}
  \draw[redo] (hr) rectangle ++(.42,.18);
\end{scope}
\end{tikzpicture}
\endgroup
\vspace{-7mm}
\caption{From no reuse toward unified ranking. (a)
Separate models restart both sides after the filter; (b) user-side reuse
restarts only the candidate side; (c) OmniPreRank's deep round
resumes from retained candidates' cached states, while
fine-ranking still restarts; (d) the same continuation at every filter
leaves one model with a shallow and a deep depth.}
\Description{Four stacked rows, each a sequence of model blocks separated
by filter lines. Each block has a purple top band for user-side tokens and
a white body for candidate-side tokens; hatched red parts start from
scratch, and green arrows across a filter mark reused computation. Row (a),
separate stage models: a pre-ranking block, a filter, and a fully hatched
fine-ranking block, so both sides start from scratch. Row (b), user-side
reuse: a green arrow carries the user band across the filter, while the
candidate body of the fine-ranking block is hatched. Row (c), OmniPreRank
(this work): shallow pre-ranking and deep pre-ranking blocks are joined
across a filter by green arrows for both the user band and the candidate
body, followed by a filter and a fully hatched fine-ranking block. Row (d),
unified ranking (outlook): Shallow and Deep blocks of one model joined by
green arrows, followed by a dashed filter, faint dashed arrows, and a
dashed block labeled deeper, if needed.}
\label{fig:reuse-concept}
\end{figure}

\looseness=-1
Recent advances in large language models (LLMs), particularly in unified architectures and large-scale training, have encouraged researchers to revisit the multi-stage organization of industrial recommendation systems. Generative recommendation, exemplified by OneRec~\cite{deng2025onerec}, directly generates recommendations with a single model, demonstrating the potential to unify retrieval and ranking end to end. Industrial advertising systems, however, have evolved over many years, tightly coupling prediction models~\cite{bian2022can,chan2022keep,chan2023hdr,chan2023pssa,sheng2024multimodal,chan2025mal,wu2026muse,liu2026est,jin2026tga,yi2026native,luo2026netcvr} with bidding, candidate filtering, business rules, and data and serving workflows. A wholesale architectural transition therefore requires coordinated adaptation and validation across all of these components. This motivates a complementary path toward unified ranking: instead of removing the stage structure, successive stages retain their decision roles while their models and computation are progressively merged. Along this path, a later stage builds on earlier computation rather than restarting from scratch.

Computation reuse offers a concrete way to pursue this path, and existing studies reuse different kinds of state.
HSTU~\cite{zhai2024hstu} uses M-FALCON to reuse user-history computation across candidate microbatches within a ranking stage.
UniSGR~\cite{sun2026unisgr} reuses semantic representations and decoder states produced during generative retrieval for subsequent multi-objective ranking.
OneTrans-V2~\cite{cao2026onetransv2} encodes candidate-\hspace{0pt}independent user context once and shares it across retrieval, pre-ranking, and fine-ranking.
As shown in Figure~\ref{fig:reuse-concept}(b), a common line of work focuses on reusing candidate-independent user computation within or across stages.
In a score-based cascade, however, candidates are filtered by ranking scores, bids, and business rules between scoring passes, and the candidate-dependent computation, which grows with the number of candidates, is typically repeated in each pass, since successive passes are served by separate models.
We therefore ask whether the model can resume computation after filtering from the intermediate states already computed for the retained candidates, rather than recomputing them, even though the filter lies outside the model.
If so, a single model can score all candidates with its first few layers and apply its remaining layers only to the candidates that pass the filter, without repeating the computation already performed for them.

Pre-ranking is a particularly suitable entry point for this question. It receives tens of thousands of retrieved candidates per request and decides which enter the more expensive fine-ranking stage~\cite{wang2020cold}. Because every candidate must be scored, any additional per-candidate computation, such as more layers or richer features, is incurred across the full candidate set, making the tension between candidate coverage and modeling depth especially acute. Directly merging pre-ranking with fine-ranking would involve coordinated changes to the features, objectives, and serving infrastructure of fine-ranking, as well as to the bidding and business logic that consume its outputs. We instead split pre-ranking into two scoring rounds within one model: a shallow round scores all candidates for a first filter, and a deep round continues from the shallow states of the retained candidates. This split addresses the cost tension within pre-ranking, and it allows the underlying mechanism, namely scoring a large candidate set, filtering by bids and business rules, and continued computation on the retained candidates, to be studied within a single stage without restructuring the entire ranking pipeline. Because retrieval, bidding policies, business rules, and fine-ranking remain unchanged, the design can be deployed and validated online. More broadly, it suggests a view of unified ranking in which the ranking cascade reduces to a shallow and a deep depth of a single model, with further depths added only when needed: were the deep round extended, beyond the scope of this paper, to take over fine-ranking, the filter between these rounds would become the boundary between pre-ranking and fine-ranking.

We instantiate this two-round design as \emph{OmniPreRank}, a unified multi-task Transformer for advertising pre-ranking. In our advertising system, candidates are ranked by effective cost per mille (ECPM), the product of a ranking signal and a bid, where the bid is computed by the bidding policy from the predicted transaction value (pGMV)~\cite{zhu2017ocpc,li2026gmvdelay}. Because bidding is performed in both rounds, OmniPreRank predicts both the ranking signal and pGMV from one shared backbone, with a set of heads at each of two depths: Shallow heads after the first few layers and Deep heads after the full backbone. Joint training supervises the early layers with both the Shallow and Deep objectives, so they serve the Shallow predictions as well as the deep computation built on top of them. At inference time, the Shallow heads score all candidates, the bidding policy converts their pGMV into bids, and the resulting ECPM scores determine the first filter. Deep resumes from retained candidates' shallow states and updates predictions for re-bidding and final filtering.

Realizing computation reuse across filtering requires model inference and the intervening business operations to be designed together. The two rounds are separated by bidding, engine-side aggregation, and business rules, with the deep round invoked through a new request. We therefore revisit the data-transfer and execution boundaries around these operations, bringing prediction and bidding closer together while preserving model states across filtering. Bidding applies bidding parameters to pGMV, but in our previous deployment it was a CPU service in a different data center, called over a conventional remote procedure call (RPC) protocol~\cite{birrell1984rpc}. Calling it in both rounds would add a cross-data-center round trip to each round, with a particularly large payload in the first round, which covers tens of thousands of candidates. We therefore run bidding on GPUs in the same data center as prediction. The bidding service processes candidates in parallel and exchanges pGMV predictions and bids with the prediction service over a low-latency connection. The model is served through separate Shallow and Deep execution graphs. The first request executes Shallow, requests bids, and caches the intermediate states required by Deep in GPU memory. The second request is routed back to the machine and GPU holding these states; Deep resumes from the states of the retained candidates, and bids are recomputed from the updated pGMV. With cached states available, retained candidates skip feature retrieval, embedding lookup, tokenization, and shared-prefix computation.

We evaluate OmniPreRank in a large-scale advertising system through offline
replay, serving benchmarks, and online A/B tests. At the deployed
$16{,}000\rightarrow2{,}000\rightarrow400$ funnel, the shallow filter
preserves nearly all of the deep output's top-400 candidates from the full
pool, while limiting deep computation to 2,000 candidates. Online,
OmniPreRank retains 91.1\% of the RPM gain of full-candidate deep scoring,
increasing RPM by 2.14\% with 36.7\% fewer prediction machines than the production
baseline under equal request load.

In summary, the main contributions of this work are threefold:
\begingroup
\settowidth{\leftmargini}{\labelitemi}
\addtolength{\leftmargini}{\labelsep}
\begin{list}{\labelitemi}{%
  \setlength{\leftmargin}{\leftmargini}%
  \setlength{\labelwidth}{\leftmargin}%
  \addtolength{\labelwidth}{-\labelsep}%
  \setlength{\topsep}{4pt}%
  \setlength{\partopsep}{0pt}%
  \setlength{\itemsep}{2pt}%
  \setlength{\parsep}{0pt}}
  \item We split pre-ranking into shallow and deep rounds of one multi-task Transformer, a step toward unified ranking through progressive computation reuse. Both depths predict the ranking signal and pGMV; Shallow scores all candidates, while Deep continues from retained candidates' intermediate states.
  \item We reorganize the serving topology to streamline the online critical path: the prediction service exchanges data directly with GPU bidding over a low-latency connection and computes ECPM, while the business engine retains cross-package aggregation and candidate filtering. Separate execution graphs, GPU state caching, and request routing let Deep resume from retained candidates' states without repeating input preparation or prefix computation.
  \item Offline and online experiments show a better trade-off between ranking quality and serving efficiency. We compare full-pool Deep scoring, two-round recomputation, and state reuse, and report bidding latency before and after the serving redesign.
\end{list}
\endgroup

\section{Preliminaries}
\label{sec:problem-definition}

\subsection{Pre-Ranking in the Advertising Cascade}
\label{subsec:candidate-selection-budgets}
We consider pre-ranking in a cascaded advertising system. For a request
$q$, retrieval returns a candidate set $\mathcal{C}_q$ of up to
$16{,}000$ advertisements, each associated with user, advertisement, and
contextual inputs. Pre-ranking scores these candidates and, together with
business rules $\mathcal{R}_q$, selects at most $K$ of them for
fine-ranking:
\begin{equation}
\mathcal{C}_q^{\mathrm{out}}=
\operatorname{Select}_{K}\!\left(
\mathcal{C}_q,\boldsymbol{s}_q;\mathcal{R}_q
\right),
\quad
\mathcal{C}_q^{\mathrm{out}}\subseteq\mathcal{C}_q,\quad
|\mathcal{C}_q^{\mathrm{out}}|\leq K,
\label{eq:candidate-selection}
\end{equation}
where $\boldsymbol{s}_q$ collects the candidates' ECPM scores. For score-only
filtering, $\operatorname{Select}$ reduces to top-$K$ selection. The production
baseline uses a single scoring round with $K=400$. The engine merges scores
and bids from all parallel scoring packages before applying
$\operatorname{Select}$ outside the model; the next model pass requires a new request.

\subsection{Scoring, Bidding, and Training Objectives}
\label{subsec:prediction-bidding-ecpm}
In our system, an ECPM score combines a learned ranking signal with a bid
to rank each candidate $i$:
\begin{equation}
s_i^{\mathrm{ECPM}}=\mathrm{pLTR}_i\,b_i,
\label{eq:ecpm-ranking-score}
\end{equation}
where $\mathrm{pLTR}_i$ is the predicted learning-to-rank (LTR) signal and $b_i$
the bid. The value branch predicts two components: $\mathrm{pCVR}_i$,
the purchase probability conditional on a click, and $\mathrm{pAMT}_i$,
the expected transaction amount conditional on a purchase.
The existing bidding policy $\mathcal{B}$ maps the predicted
transaction value to a bid:
\begin{equation}
b_i=\mathcal{B}\!\left(\mathrm{pGMV}_i;\boldsymbol{z}_i\right),
\label{eq:value-dependent-bidding}
\end{equation}
where $\mathrm{pGMV}_i=\mathrm{pCVR}_i\,\mathrm{pAMT}_i$ is the predicted
transaction value conditional on a click, and $\boldsymbol{z}_i$ denotes
the bidding parameters associated with the advertiser, candidate, and
request.

Training uses candidates from fine-ranking logs. Following COPR's rank
alignment~\cite{zhao2023copr}, sampled candidates from each request's
fine-ranking ECPM list form a pair set $\mathcal{P}$, with
$(i,j)\in\mathcal{P}$ when $i$ ranks above $j$. The LTR loss encourages
pre-ranking ECPM to preserve this order:
\begin{equation}
\mathcal{L}_{\mathrm{LTR}}=
\sum_{(i,j)\in\mathcal{P}}
\log\!\Big[1+\exp\!\Big(-\Big(
\frac{\mathrm{pLTR}_i\,b_i}{\mathrm{pLTR}_j\,b_j}-1\Big)\Big)\Big]
+R,
\label{eq:ltr-loss}
\end{equation}
where $R$ regularizes a per-candidate relaxation factor in
$\mathrm{pLTR}_i$ (Appendix~\ref{subsec:appendix-model-training}). Here,
$b_i$ is the logged pre-ranking bid, so this loss is independent of
predicted pGMV. Fine-ranking ECPM orders, rather than click labels,
supervise $\mathrm{pLTR}_i$ as a ranking signal, not a calibrated
click-through probability.

The value branch distills fine-ranking predictions for pCVR and pAMT.
Let $\mathcal{K}=\{\mathrm{CVR},\mathrm{AMT}\}$, with
$\widehat{y}_{i,\mathrm{CVR}}=\mathrm{pCVR}_i$ and
$\widehat{y}_{i,\mathrm{AMT}}=\mathrm{pAMT}_i$. For the fine-ranking
training set $\mathcal{I}$, we compute mean squared error (MSE) after
transforming both predictions $\widehat{y}_{i,k}$ and fine-ranking
targets $\widetilde{y}_{i,k}^{\mathrm{F}}$ with
$g_{\mathrm{CVR}}(p)=\log\frac{p}{1-p}$ or
$g_{\mathrm{AMT}}(x)=\log(1+x)$:
\begin{equation}
\mathcal{L}_{\mathrm{V}}=
\sum_{k\in\mathcal{K}}\frac{1}{|\mathcal{I}|}\sum_{i\in\mathcal{I}}
\left[g_k\!\left(\widehat{y}_{i,k}\right)
-g_k\!\left(\widetilde{y}_{i,k}^{\mathrm{F}}\right)\right]^2.
\label{eq:value-loss}
\end{equation}
Appendix~\ref{subsec:appendix-model-training} provides the remaining
model and training details.

\subsection{Serving Cost}
\label{subsec:serving-cost}
Pre-ranking must meet latency and throughput requirements under finite
GPU compute and memory. We use a simplified model of candidate-dependent
processing cost. Consider a pre-ranking model with $L$ layers whose
candidate-dependent computation is independent across candidates. For each candidate, let $c_{\mathrm{in}}$ denote the cost of input
preparation, including feature retrieval, embedding lookup, and
tokenization; $c_\ell$ the cost of layer $\ell$; and $c_{\mathrm{bid}}$
the cost of bidding. Candidate-independent work, such as user-side
computation shared across candidates (the top band in
Figure~\ref{fig:reuse-concept}), is not multiplied by the number of
candidates and is omitted. The per-request cost of single-round
pre-ranking is then
\begin{equation}
\mathrm{Cost}_{\mathrm{single}}=
|\mathcal{C}_q|\Big(c_{\mathrm{in}}+\sum_{\ell=1}^{L}c_\ell
+c_{\mathrm{bid}}\Big).
\label{eq:single-round-cost}
\end{equation}
Each component is incurred across the full candidate set: increasing
modeling depth or adding more expensive candidate-dependent features
raises the cost for all candidates. This model captures
candidate-dependent processing work rather than end-to-end latency, which
also depends on parallel execution, memory traffic, and communication
overhead. The challenge is to apply deeper computation to fewer candidates
without degrading selection: candidates filtered out early cannot be
recovered downstream.

\section{OmniPreRank}
\label{sec:method}

\begin{figure*}[t]
\centering
\resizebox{\textwidth}{!}{\input{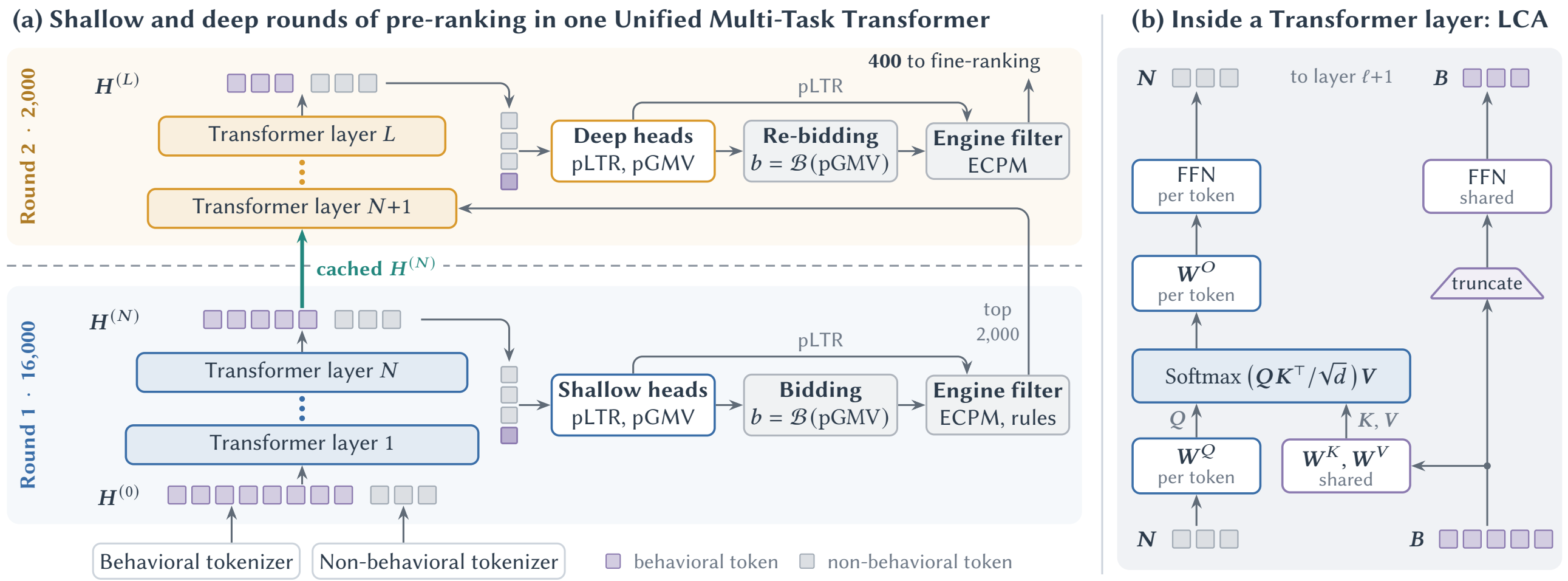}}
\caption{OmniPreRank overview. (a) Round~1 scores all candidates with the
first $N$ layers and Shallow heads; after bidding and engine filtering,
round~2 resumes the retained candidates' cached layer-$N$ states through the
remaining layers and Deep heads. Heads read non-behavioral tokens
and the mean behavioral token (darker square). Both depths train on every example;
filtering is inference-only. (b) LCA: non-behavioral tokens query
behavioral tokens, which are truncated per layer before the shared FFN
(pre-normalization and residual connections omitted for clarity).}
\Description{Panel (a) shows stacked Transformer layers. Behavioral and
non-behavioral tokenizers feed layers 1 to N for all 16,000 candidates.
The non-behavioral tokens and the mean behavioral token are read out by
the Shallow heads, followed by bidding and engine filtering. A dashed line
separates the two rounds, which arrive as separate requests. The cached
states of the 2,000 selected candidates continue through layers N+1 to L,
whose readout feeds the Deep heads, re-bidding, and engine filtering that
sends 400 candidates to fine-ranking. Panel (b) shows one layer:
non-behavioral tokens pass through per-token query projections, behavioral
tokens through shared key and value projections, and attention combines
them before a per-token output projection; behavioral tokens are updated
and truncated separately.}
\label{fig:overview}
\end{figure*}

Figure~\ref{fig:overview} overviews OmniPreRank: one multi-task Transformer serves both rounds through Shallow and Deep heads at two depths.

\subsection{Unified Multi-Task Transformer}
\label{subsec:unified-multi-task-transformer}
The backbone builds on EST~\cite{liu2026est}, adopting its tokenization,
layer design, and prediction readout. Non-behavioral features, including
user, advertisement, and context attributes, are mapped to tokens of a
common dimension $d$, and behavioral sequences are represented as tokens in
the same space. Following EST, the behavioral sequences comprise
user-specific sequences, such as recent behaviors, and candidate-specific
sequences that a general search unit (GSU) retrieves from the user's
lifelong behaviors for each candidate~\cite{liu2026est,wu2026muse}. Features from the separate ranking-signal and pGMV models
enter one backbone serving both tasks.

Each layer follows EST rather than a standard Transformer layer, which
would apply full self-attention over all tokens. We write the states of
candidate $i$ after layer $\ell$ as non-behavioral tokens
$\boldsymbol{N}_i^{(\ell)}\in\mathbb{R}^{n_{\mathrm{N}}\times d}$ and
behavioral tokens
$\boldsymbol{B}_i^{(\ell)}\in\mathbb{R}^{n_{\mathrm{B}}^{(\ell)}\times d}$.
Lightweight cross-attention (LCA) uses non-behavioral queries and
behavioral keys and values; omitting layer and candidate indices and writing
single-head attention for brevity,
\begin{gather}
(\boldsymbol{Q}_{\boldsymbol{N}},\boldsymbol{K}_{\boldsymbol{B}},
\boldsymbol{V}_{\boldsymbol{B}})
=\big(\phi(\boldsymbol{N},\boldsymbol{W}^{Q}),\
\boldsymbol{B}\boldsymbol{W}^{K},\ \boldsymbol{B}\boldsymbol{W}^{V}\big),
\label{eq:lca-projection}\\
\operatorname{LCA}(\boldsymbol{N},\boldsymbol{B})
=\phi\!\left(\operatorname{Softmax}\!\left(
\frac{\boldsymbol{Q}_{\boldsymbol{N}}\boldsymbol{K}_{\boldsymbol{B}}^{\top}}
{\sqrt{d}}\right)\boldsymbol{V}_{\boldsymbol{B}},\
\boldsymbol{W}^{O}\right),
\label{eq:lca}
\end{gather}
where $\phi(\boldsymbol{N},\boldsymbol{W})$ multiplies the $j$-th
non-behavioral token by its own matrix
$\boldsymbol{W}_j\in\mathbb{R}^{d\times d}$, so $\boldsymbol{W}^{Q}$ and
$\boldsymbol{W}^{O}$ hold one projection per non-behavioral token, whereas
$\boldsymbol{W}^{K},\boldsymbol{W}^{V}\in\mathbb{R}^{d\times d}$ are shared
by the tokens of each behavioral sequence, with separate parameters per
sequence. Each layer applies pre-normalization and residual connections;
after LCA, a per-token FFN updates the non-behavioral tokens, while behavioral
tokens are truncated and updated by an FFN shared within each sequence. Because behavioral tokens
serve only as keys and values in LCA, its attention costs $O(n_{\mathrm{N}}n_{\mathrm{B}}d)$ rather than the
$O\big((n_{\mathrm{N}}+n_{\mathrm{B}})^2d\big)$ of full self-attention.
User-specific sequences do not depend on the
candidate, so their keys and values are computed once per package and
shared by its candidates; candidate-specific sequences are processed for
each candidate.

For candidate $i$, let $\boldsymbol{H}_i^{(0)}$ denote the tokenized input
and $\boldsymbol{H}_i^{(\ell)}$ the hidden token states after backbone
layer $\ell$:
\begin{equation}
\boldsymbol{H}_i^{(\ell)}
=\mathcal{F}_{\ell}\!\left(\boldsymbol{H}_i^{(\ell-1)}\right),
\qquad \ell=1,\ldots,L,
\label{eq:shared-backbone}
\end{equation}
where $L=N+M$ and $\mathcal{F}_{\ell}$ includes layer computation and
configured sequence truncation. Following OneTrans~\cite{zhang2025onetrans},
$n_{\mathrm{B}}^{(\ell)}$ decreases with depth while $n_{\mathrm{N}}$
is fixed, so the prefix processes the longest sequences. Token truncation
acts within each candidate, while inference-time filtering selects
across candidates. Apart from shared user-specific keys and values,
candidate forward computations are independent, enabling continuation
after filtering (Section~\ref{subsec:two-stage-inference}).

At prediction depth $\ell$, we mean-pool behavioral tokens and concatenate
the result with non-behavioral tokens:
\begin{equation}
\boldsymbol{r}_i^{(\ell)}=
\operatorname{Concat}\!\left(
\operatorname{vec}(\boldsymbol{N}_i^{(\ell)}),
\operatorname{Mean}(\boldsymbol{B}_i^{(\ell)})
\right).
\label{eq:prediction-readout}
\end{equation}

\subsection{Shallow--Deep Heads and Joint Training}
\label{subsec:joint-shallow-deep-training}
Prediction heads are attached at two depths. Let
$a\in\{\mathrm{S},\mathrm{D}\}$ index Shallow and Deep, with prediction
depths $\ell_{\mathrm{S}}=N$ and $\ell_{\mathrm{D}}=L$, which serve the first
and second rounds, respectively. Both rounds bid, so each readout predicts
the ranking signal and the two value components of
Section~\ref{subsec:prediction-bidding-ecpm}:
\begin{equation}
\boldsymbol{p}_i^{a}
=h_a\!\left(\boldsymbol{r}_i^{(\ell_a)}\right)
=\left(\mathrm{pLTR}_i^{a},\,\mathrm{pCVR}_i^{a},\,\mathrm{pAMT}_i^{a}\right),
\label{eq:shallow-deep-outputs}
\end{equation}
where $\mathrm{pLTR}_i^{a}$ enters Eq.~\eqref{eq:ecpm-ranking-score} and
the composed value $\mathrm{pGMV}_i^{a}=\mathrm{pCVR}_i^{a}\,\mathrm{pAMT}_i^{a}$
enters Eq.~\eqref{eq:value-dependent-bidding}. The
head sets $h_{\mathrm{S}}$ and $h_{\mathrm{D}}$ have independent
parameters and share the feature representations and prefix computation
through the backbone. The configurable split position is $N=1$ in deployment, so the
shallow round runs one backbone layer on all candidates.

Training runs the full backbone on every example and applies the
objectives of Section~\ref{subsec:prediction-bidding-ecpm} at both depths.
Let $\mathcal{L}_{\mathrm{LTR}}^{a}$ and $\mathcal{L}_{\mathrm{V}}^{a}$
denote the depth-$a$ losses in Eqs.~\eqref{eq:ltr-loss}
and~\eqref{eq:value-loss}. The joint objective is
\begin{equation}
\mathcal{L}
=\sum_{a\in\{\mathrm{S},\mathrm{D}\}}
\left(\lambda_a^{\mathrm{LTR}}\mathcal{L}_{\mathrm{LTR}}^{a}
+\lambda_a^{\mathrm{V}}\mathcal{L}_{\mathrm{V}}^{a}\right),
\label{eq:joint-training}
\end{equation}
with fixed weights. Deep losses backpropagate through the suffix into the
shared prefix, while Shallow losses supervise the prefix directly, so the
early layers serve both the Shallow predictions and the deep computation
built on them.

At inference
time, the shallow round scores the full retrieval set, a broader
distribution than its training data, a gap shared by pre-ranking models
trained on fine-ranking candidates. The deep round scores only the $K_1$
retained candidates, which are closer to the candidates it is trained on.

\subsection{Two-Round Inference with Shallow-to-Deep Continuation}
\label{subsec:two-stage-inference}
OmniPreRank replaces the single scoring round of
Section~\ref{subsec:candidate-selection-budgets} with two rounds. Let
$\mathcal{C}_q^{(0)}=\mathcal{C}_q$. The first round applies
Eq.~\eqref{eq:candidate-selection} with its own ECPM scores and budget.
Business rules apply only in round~1; round~2 keeps the top $K_2$
retained candidates by ECPM:
\begin{equation}
\begin{aligned}
\mathcal{C}_q^{(1)} &=
\operatorname{Select}_{K_1}\!\left(
\mathcal{C}_q^{(0)},\boldsymbol{s}_q^{(1)};\mathcal{R}_q
\right), \\
\mathcal{C}_q^{(2)} &=
\operatorname{TopK}_{K_2}\!\left(
\mathcal{C}_q^{(1)},\boldsymbol{s}_q^{(2)}
\right),
\end{aligned}
\label{eq:two-round-selection}
\end{equation}
where $\boldsymbol{s}_q^{(1)}$ and $\boldsymbol{s}_q^{(2)}$ are computed
through Eqs.~\eqref{eq:ecpm-ranking-score}
and~\eqref{eq:value-dependent-bidding} from the Shallow and Deep outputs,
respectively, and $K_2=K$. The deployed funnel is
$16{,}000\rightarrow2{,}000\rightarrow400$, so the final output size
matches the single-round system's budget for fine-ranking.

In the shallow round, every candidate in $\mathcal{C}_q^{(0)}$ goes
through input preparation and the first $N$ layers. The Shallow heads
produce $\mathrm{pLTR}_i^{\mathrm{S}}$ and the value components
$\mathrm{pCVR}_i^{\mathrm{S}},\mathrm{pAMT}_i^{\mathrm{S}}$; the bidding
policy converts their product $\mathrm{pGMV}_i^{\mathrm{S}}$ into bids, and
Eq.~\eqref{eq:two-round-selection} with budget $K_1$ yields
$\mathcal{C}_q^{(1)}$. The hidden states $\boldsymbol{H}_i^{(N)}$ at the
prefix--suffix boundary are retained. They include all non-behavioral and
behavioral states required by the suffix, including those of the
candidate-specific sequences, whereas the pooled readout of
Eq.~\eqref{eq:prediction-readout} serves only the Shallow heads.

The deep round accesses the retained states by the selected candidate
indices and executes the remaining $M$ layers:
\begin{equation}
\boldsymbol{H}_i^{(L)}
=\mathcal{F}_{\mathrm{post}}\!\left(\boldsymbol{H}_i^{(N)}\right),
\qquad i\in\mathcal{C}_q^{(1)},
\label{eq:deep-continuation}
\end{equation}
where $\mathcal{F}_{\mathrm{post}}=\mathcal{F}_{L}\circ\cdots\circ\mathcal{F}_{N+1}$.
Because candidates do not interact in the forward computation
(Section~\ref{subsec:unified-multi-task-transformer}), each retained
candidate obtains the same layer-$L$ states as it would by running the
full backbone. The deep round therefore evaluates the same function on
which the Deep heads are trained, without approximation. The Deep heads
produce $\mathrm{pLTR}_i^{\mathrm{D}}$ and the value components
$\mathrm{pCVR}_i^{\mathrm{D}},\mathrm{pAMT}_i^{\mathrm{D}}$; bids are
recomputed from their product $\mathrm{pGMV}_i^{\mathrm{D}}$ with cached
parameters $\boldsymbol{z}_i$, and
Eq.~\eqref{eq:two-round-selection} with budget $K_2$ yields
$\mathcal{C}_q^{(2)}$ for fine-ranking.

\subsection{Cost Analysis}
\label{subsec:cost-analysis}
With $L=N+M$, Eq.~\eqref{eq:single-round-cost} gives the two-round cost
per request:
\begin{equation}
\begin{aligned}
\mathrm{Cost}_{\mathrm{two}}
&= |\mathcal{C}_q^{(0)}|\Big(c_{\mathrm{in}}+\sum_{\ell=1}^{N}c_\ell
+c_{\mathrm{bid}}\Big) \\
&\quad +|\mathcal{C}_q^{(1)}|\Big(\sum_{\ell=N+1}^{L}c_\ell+c_{\mathrm{bid}}\Big).
\end{aligned}
\label{eq:two-round-cost}
\end{equation}
Two references separate the sources of savings. Compared with applying
all $L$ layers to every candidate in a single round, the two-round design
saves $(|\mathcal{C}_q^{(0)}|-|\mathcal{C}_q^{(1)}|)\sum_{\ell>N}c_\ell$
at the cost of a second bidding pass over $\mathcal{C}_q^{(1)}$; this is
the saving from candidate reduction. Compared with a two-round design
whose deep round recomputes from the inputs, continuation removes
$|\mathcal{C}_q^{(1)}|\big(c_{\mathrm{in}}+\sum_{\ell\le N}c_\ell\big)$;
this is the saving from state reuse. Because behavioral sequences,
including the candidate-specific ones, are truncated with depth
(Section~\ref{subsec:unified-multi-task-transformer}), the prefix layers
process the longest sequences, so continuation removes more work than the
prefix's share of layers suggests. The expression omits the cost of writing,
storing, and reading the retained states, which
Section~\ref{sec:online-deployment} addresses.

\section{Online Deployment}
\label{sec:online-deployment}
The computation reuse in Section~\ref{subsec:two-stage-inference} assumes that states retained by the shallow round remain available to the deep round. In online serving, this reuse requires coordinating model inference with the intervening business operations: bidding consumes model predictions, while engine-side aggregation and business rules determine which candidates return for deeper computation in a separate request routed back to the GPUs holding their states.

The deployment therefore addresses two linked requirements: low-latency data exchange between prediction and GPU bidding, and reliable preservation and retrieval of candidate states across filtering and requests. A direct connection between the two services addresses the first; GPU state caching and request routing address the second. The following subsections describe the workflow and these mechanisms, including cache management and fallback.

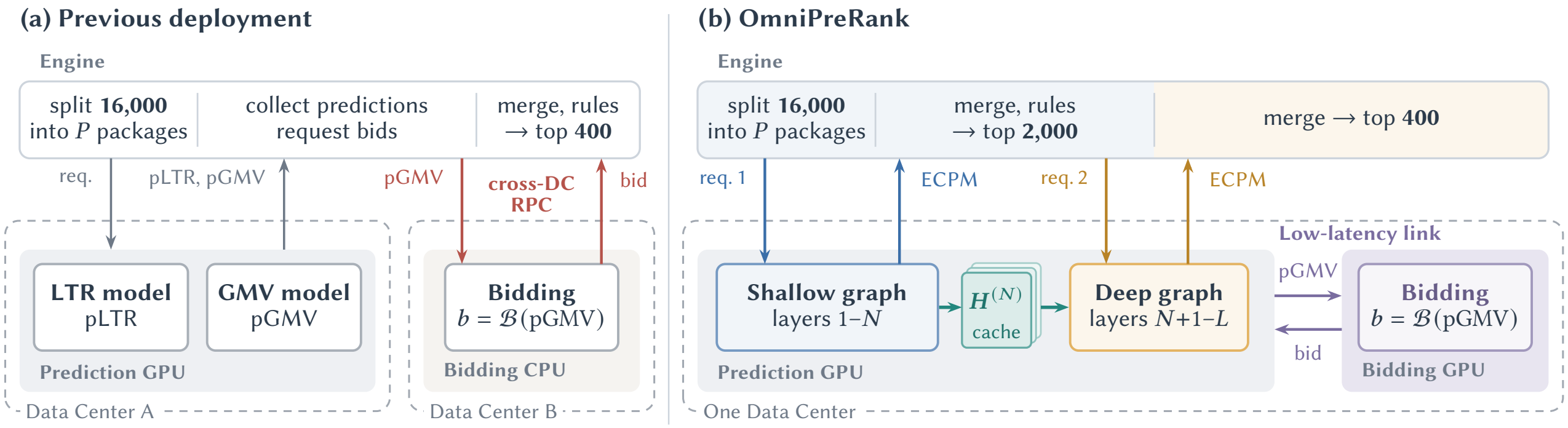
\begin{figure*}[t]
\centering
\resizebox{\textwidth}{!}{
\begingroup
\definecolor{omniInk}{HTML}{2B3A4A}
\definecolor{omniMuted}{HTML}{6E7987}
\definecolor{omniLine}{HTML}{CDD4DC}
\definecolor{omniShallow}{HTML}{3A6EA8}
\definecolor{omniDeep}{HTML}{D99A2E}
\definecolor{omniState}{HTML}{23877E}
\definecolor{omniBidding}{HTML}{796C9F}
\definecolor{omniRpc}{HTML}{B5524A}
\definecolor{omniCpuBg}{HTML}{F5F3F0}
\definecolor{omniGpuBg}{HTML}{E8E5F0}
\begin{tikzpicture}[
  x=1cm,y=1cm,
  font=\sffamily\fontsize{7.2}{8.6}\selectfont,
  text=omniInk,
  >={Stealth[length=1.5mm,width=1.1mm]},
  enginebar/.style={fill=white,draw=omniLine,line width=.6pt,
    rounded corners=3pt},
  card/.style={fill=omniLine!35,rounded corners=3pt},
  dc/.style={fill=white,draw=omniMuted!65,
    dash pattern=on 2.4pt off 2pt,line width=.55pt,rounded corners=4pt},
  unit/.style={rounded corners=2.5pt,line width=.7pt,align=center,
    inner sep=2pt,minimum height=.9cm},
  neutral/.style={unit,fill=white,draw=omniMuted!60},
  shallow/.style={unit,fill=omniShallow!7,draw=omniShallow!70},
  deep/.style={unit,fill=omniDeep!9,draw=omniDeep!75},
  flow/.style={->,draw=omniMuted,line width=.65pt},
  rpc/.style={->,draw=omniRpc,line width=.8pt},
  biddinglink/.style={->,draw=omniBidding,line width=.8pt},
  state/.style={->,draw=omniState,line width=1pt},
  paneltitle/.style={font=\sffamily\bfseries\fontsize{7.8}{9.4}\selectfont,
    text=omniInk,anchor=base west,inner sep=0pt},
  label/.style={font=\sffamily\bfseries\fontsize{6.1}{7.2}\selectfont,
    text=omniMuted,anchor=west,inner sep=0pt},
  note/.style={font=\sffamily\fontsize{6.4}{7.5}\selectfont,
    text=omniMuted,inner sep=1pt,align=center},
  dclabel/.style={note,anchor=west,fill=white,inner sep=1.5pt},
  link/.style={font=\sffamily\bfseries\fontsize{6.3}{7.4}\selectfont,
    align=center,inner sep=1pt}
]
\path[use as bounding box] (-.02,.06) rectangle (16.32,4.81);

\node[paneltitle] at (.20,4.42) {(a) Previous deployment};
\node[label] at (.40,4.04) {Engine};

\path[enginebar] (.20,3.05) rectangle (6.65,3.85);
\draw[omniLine,line width=.5pt] (2.05,3.17) -- (2.05,3.73);
\draw[omniLine,line width=.5pt] (4.95,3.17) -- (4.95,3.73);
\node[align=center] at (1.125,3.45)
  {split \textbf{16,000}\\[-1pt]into $P$ packages};
\node[align=center] at (3.50,3.45)
  {collect predictions\\[-1pt]request bids};
\node[align=center] at (5.80,3.45)
  {merge, rules\\[-1pt]$\rightarrow$ top \textbf{400}};

\path[dc] (.05,.42) rectangle (4.05,2.40);
\path[dc] (4.25,.42) rectangle (6.80,2.40);
\path[card] (.20,.62) rectangle (3.90,2.10);
\path[card,fill=omniCpuBg] (4.40,.62) rectangle (6.65,2.10);
\node[label] at (.40,.83) {Prediction GPU};
\node[label] at (4.60,.83) {Bidding CPU};
\node[dclabel] at (.20,.42) {Data Center A};
\node[dclabel] at (4.40,.42) {Data Center B};

\node[neutral,minimum width=1.6cm] (altr) at (1.15,1.5)
  {\textbf{LTR model}\\[-1pt]pLTR};
\node[neutral,minimum width=1.6cm] (agmv) at (2.95,1.5)
  {\textbf{GMV model}\\[-1pt]pGMV};
\node[neutral,minimum width=1.8cm] (abid) at (5.525,1.5)
  {\textbf{Bidding}\\[-1pt]$b=\mathcal{B}(\mathrm{pGMV})$};

\draw[flow] (1.15,3.05) -- (1.15,2.10);
\node[note,anchor=east] at (1.00,2.83) {req.};
\draw[flow] (2.95,2.10) -- (2.95,3.05);
\node[note,anchor=east] at (2.80,2.83) {pLTR, pGMV};
\draw[rpc] (4.80,3.05) -- (4.80,1.95);
\node[note,text=omniRpc,anchor=east] at (4.65,2.83) {pGMV};
\draw[rpc] (6.25,1.95) -- (6.25,3.05);
\node[note,text=omniRpc,anchor=west] at (6.40,2.83) {bid};
\node[link,text=omniRpc] at (5.525,2.68) {cross-DC\\[-1pt]RPC};

\draw[omniLine,line width=.45pt] (6.95,.28) -- (6.95,4.55);

\begin{scope}[xshift=-1.7cm]
\node[paneltitle] at (8.95,4.42) {(b) OmniPreRank};
\node[label] at (9.15,4.04) {Engine};

\begin{scope}
  \clip[rounded corners=3pt] (8.95,3.05) rectangle (17.8,3.85);
  \fill[omniShallow!7] (8.95,3.05) rectangle (13.7,3.85);
  \fill[omniDeep!9] (13.7,3.05) rectangle (17.8,3.85);
\end{scope}
\draw[draw=omniLine,line width=.6pt,rounded corners=3pt]
  (8.95,3.05) rectangle (17.8,3.85);
\draw[omniLine,line width=.5pt] (10.8,3.17) -- (10.8,3.73);
\draw[omniLine,line width=.5pt] (13.7,3.17) -- (13.7,3.73);
\node[align=center] at (9.875,3.45)
  {split \textbf{16,000}\\[-1pt]into $P$ packages};
\node[align=center] at (12.25,3.45)
  {merge, rules\\[-1pt]$\rightarrow$ top \textbf{2,000}};
\node[align=center] at (15.75,3.45)
  {merge $\rightarrow$ top \textbf{400}};

\path[dc] (8.80,.42) rectangle (17.95,2.40);
\path[card] (8.95,.62) rectangle (14.95,2.10);
\path[card,fill=omniGpuBg] (15.65,.62) rectangle (17.8,2.10);
\node[label] at (9.15,.83) {Prediction GPU};
\node[label] at (15.85,.83) {Bidding GPU};
\node[dclabel] at (8.95,.42) {One Data Center};

\node[shallow,minimum width=2.3cm] (shallow) at (10.3,1.5)
  {\textbf{Shallow graph}\\[-1pt]layers $1$--$N$};

\fill[omniState!7,draw=omniState!40,line width=.5pt,rounded corners=1.5pt]
  (11.80,1.18) rectangle (12.52,1.96);
\fill[omniState!11,draw=omniState!50,line width=.5pt,rounded corners=1.5pt]
  (11.75,1.13) rectangle (12.47,1.91);
\fill[omniState!16,draw=omniState!70,line width=.6pt,rounded corners=1.5pt]
  (11.70,1.08) rectangle (12.42,1.86);
\node[text=omniState!85!black] at (12.06,1.60) {$\boldsymbol H^{(N)}$};
\node[note,text=omniState!85!black,
  font=\sffamily\fontsize{6.1}{7.2}\selectfont] at (12.06,1.25) {cache};

\node[deep,minimum width=1.85cm] (deep) at (13.75,1.5)
  {\textbf{Deep graph}\\[-1pt]layers $N{+}1$--$L$};
\node[unit,fill=omniBidding!6,draw=omniBidding!70,minimum width=1.8cm]
  (bbid) at (16.725,1.5)
  {\textbf{\textcolor{omniBidding!85!black}{Bidding}}\\[-1pt]
    $b=\mathcal{B}(\mathrm{pGMV})$};
\draw[state] (shallow.east) -- (11.70,1.5);
\draw[state] (12.52,1.5) -- (deep.west);

\draw[->,draw=omniShallow,line width=.8pt] (9.65,3.05) -- (9.65,1.95);
\node[note,text=omniShallow,anchor=east] at (9.50,2.83) {req.\,1};
\draw[->,draw=omniShallow,line width=.8pt] (11.05,1.95) -- (11.05,3.05);
\node[note,text=omniShallow,anchor=west] at (11.23,2.83) {ECPM};
\draw[->,draw=omniDeep!85!black,line width=.8pt] (13.20,3.05) -- (13.20,1.95);
\node[note,text=omniDeep!85!black,anchor=east] at (13.05,2.83) {req.\,2};
\draw[->,draw=omniDeep!85!black,line width=.8pt] (14.05,1.95) -- (14.05,3.05);
\node[note,text=omniDeep!85!black,anchor=west] at (14.23,2.83) {ECPM};

\node[link,text=omniBidding,anchor=west] at (14.95,2.25) {Low-latency link};
\draw[biddinglink] (14.95,1.62) -- (15.65,1.62);
\node[note,text=omniBidding] at (15.3,1.86) {pGMV};
\draw[biddinglink] (15.65,1.27) -- (14.95,1.27);
\node[note,text=omniBidding] at (15.3,1.03) {bid};
\end{scope}
\end{tikzpicture}
\endgroup}
\caption{Previous deployment and current OmniPreRank deployment.
(a) The engine dispatches packages to separate LTR and GMV models and calls CPU bidding across data centers.
(b) Deep resumes from cached Shallow states after engine filtering;
in both rounds, the Shallow and Deep prediction graphs directly invoke GPU bidding over a low-latency connection.}
\Description{Two panels compare the previous and current deployments. Panel (a), previous
deployment: the engine splits 16,000 candidates into P packages; in a
single round, a separate LTR model and a separate GMV model on prediction
GPUs in data center A score all candidates and return their predictions
to the engine. The engine forwards pGMV to a CPU bidding service in data
center B over cross-data-center RPC;
bids return to the engine, which merges the packages, combines pLTR and bids into ECPM, applies
rules, and sends the top 400 to fine-ranking. Panel (b), OmniPreRank:
within one data center, request 1 runs the Shallow
graph, layers 1 to N, on a package and writes the layer-N states to a GPU
cache; the engine merges the packages, applies rules, and keeps the top
2,000. Request 2 runs the Deep graph, layers N+1 to L, from the cached
states on the same GPU, and the engine merges the results and sends the
top 400 to fine-ranking. In both rounds, the prediction GPU exchanges pGMV
and bids directly with the GPU bidding service over a low-latency connection.}
\label{fig:serving-workflow}
\end{figure*}

\subsection{Two-Request Serving Workflow}
\label{subsec:two-stage-serving}
Engine-side filtering separates Shallow and Deep into two requests.
The Shallow graph prepares inputs and runs the first $N$ layers and
Shallow heads; the Deep graph resumes from cached states of retained
candidates and runs the remaining layers and Deep heads. Each graph
produces both the ranking signal and pGMV in one prediction call per
package.

Figure~\ref{fig:serving-workflow} shows the workflow. In the first request,
the engine splits up to 16,000 candidates into $P$ packages of up to
1,024 candidates ($P\leq16$) and dispatches them to GPU serving machines. Each machine runs the Shallow
graph on its package, caches the layer-$N$ states in GPU memory, requests
bids for its candidates from the bidding service, computes ECPM scores,
and returns them together with a cache identifier and the location of the
cached states. The engine aggregates the package results, applies
business rules, and selects $K_1$ candidates for the deep round.

In the second request, the engine groups the retained candidates by their
original packages and sends each group,
with its cache identifier and the candidates' original indices, back to
the machine and GPU holding the corresponding states. Each machine runs
the Deep graph from these states, requests new bids from the updated
pGMV, and returns second-round ECPM scores, from which the engine makes
the final selection. With identical operator optimizations and
mixed-precision settings, resumed execution numerically matches a full
forward pass, as verified in deployment
(Section~\ref{subsec:two-stage-inference}).

\subsection{Integrating Prediction and Bidding}
\label{subsec:same-unit-bidding}
Bidding converts the model's pGMV predictions into bids used to compute
ECPM scores for candidate selection. It lies on the critical path of both
rounds. In the previous deployment, the engine collected pGMV from the
GMV prediction service and invoked a CPU bidding service in another data
center through RPC~\cite{birrell1984rpc}. The returned bids were combined
with pLTR in the engine to compute ECPM. Retaining this path in both rounds
would repeat the engine relay and cross-data-center round trip, with a
particularly large payload for the first round's tens of thousands of
candidates.

The current deployment integrates prediction and bidding within each round's
execution path. The prediction service directly requests bids and combines
them with its ranking predictions to compute ECPM before returning package
results to the engine. The engine aggregates scores across packages,
applies business rules in the first round, and selects candidates for
further computation.

Prediction and bidding run on GPUs in the same data center. Bidding runs
in parallel across candidates, and the two services exchange pGMV
predictions and bids over a low-latency connection. The bidding policy
and its parameters remain unchanged.
Parameters are read from a local index in round~1 and cached for round~2;
caching these few coefficients adds negligible memory.

\subsection{GPU State Caching and Request Routing}
\label{subsec:gpu-state-caching-routing}
Continuation across engine-side filtering requires both locating the
cached states and matching them to the retained candidates. Each
package's layer-$N$ states are cached on the GPU that produced them under
a cache identifier. Because caching precedes the first selection, all
candidate rows in the package are stored. The engine routes the second
request to that GPU, and the cache identifier and original candidate
indices locate the package and select its retained rows. This mapping is
preserved through sorting and filtering.

Correct continuation also requires consistent model versions and safe
state lifetimes. Before a model update, the machine stops accepting
first-round requests and waits for existing cache entries to expire and
in-flight Deep computations to finish before activating the new version.
Cache entries being read by the Deep graph are protected from
reclamation, including after TTL expiry.

\subsection{Cache Capacity and Fallback}
\label{subsec:cache-capacity-fallback}
State retention across requests must fit within the GPU memory budget.
The cached states occupy about 24~KB per candidate, about 24.6~MB for a
maximum production package of 1,024 candidates and about 384~MB for a full request
of 16,000 candidates.
The cache uses a time-to-live (TTL) of 100~ms, twice the 50~ms end-to-end
deadline for the two-round pre-ranking workflow. At the maximum
sustainable load in Table~\ref{tab:serving-breakdown} (1,950 packages/s per GPU),
taking the maximum package size of 1,024 candidates as an upper bound yields
about 2.0 million candidates per second per GPU, corresponding to a resident cache
of at most 4.8~GB ($2.0\times10^{6}/\mathrm{s}\times0.1\,\mathrm{s}\times24\,\mathrm{KB}$),
or about 6.7\% of the 72~GB memory of our inference GPU. Cache writes require
at most 48~GB/s, or 3.6\% of the GPU's 1,345~GB/s memory bandwidth.

The cache has a hard capacity of 15~GB per GPU, about 3.1 times the
estimated resident size, to provide headroom for bursts. Writes that
would exceed this capacity are rejected without allocating memory,
preventing cache-induced out-of-memory failures.

Fallback operates at package granularity. If a cache write is rejected,
the machine holding the states fails, an entry expires, or the request
times out, the engine uses the affected package's Shallow ECPM scores
from round~1 for the final selection. These scores are merged directly
with Deep ECPM scores from the other packages, without cross-depth
calibration. This fallback lets the engine complete final selection
when some packages lack Deep scores.

\section{Experiments}
\label{sec:experiments}

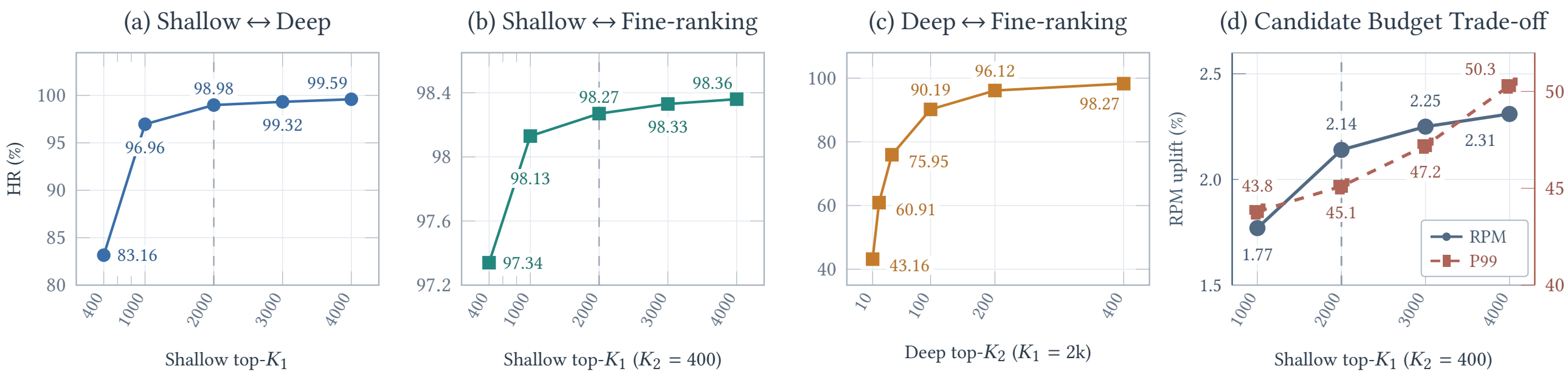
\begin{figure*}[!t]
\centering
\definecolor{coverageBlue}{HTML}{3A6EA8}
\definecolor{coverageTeal}{HTML}{23877E}
\definecolor{coverageAmber}{HTML}{C47C2C}
\definecolor{coverageInk}{HTML}{2B3A4A}
\definecolor{coverageMuted}{HTML}{667587}
\definecolor{coverageFrame}{HTML}{AAB5C1}
\definecolor{coverageGrid}{HTML}{E5EAF0}
\definecolor{budgetRPM}{HTML}{526A83}
\definecolor{budgetLatency}{HTML}{AF6458}
\definecolor{budgetInk}{HTML}{2B3A4A}
\definecolor{budgetMuted}{HTML}{667587}
\definecolor{budgetFrame}{HTML}{AAB5C1}
\definecolor{budgetGrid}{HTML}{E5EAF0}
\begin{tikzpicture}
\begin{groupplot}[
  group style={group name=coverage, group size=4 by 1, horizontal sep=0.9cm},
  scale only axis=true, width=0.185\textwidth, height=0.14210526\textwidth,
  xmode=normal, xmin=0, xmax=4400,
  xtick={400,1000,2000,3000,4000},
  xticklabels={400,1000,2000,3000,4000},
  minor xtick={600,800}, xlabel={Shallow top-$K_1$},
  xticklabel style={rotate=60, anchor=east, font=\scriptsize},
  tick label style={font=\scriptsize, text=coverageMuted}, label style={font=\scriptsize, text=coverageInk},
  title style={font=\small, text=coverageInk, yshift=-1ex},
  ylabel style={yshift=-0.6ex},
  axis line style={draw=coverageFrame, line width=0.45pt},
  tick style={draw=coverageFrame},
  grid=major, grid style={draw=coverageGrid, line width=0.3pt},
  every axis plot/.append style={mark size=1.7pt, line width=0.8pt}]
\nextgroupplot[title={(a) Shallow\,$\leftrightarrow$\,Deep},
  ylabel={HR (\%)}, ymin=80, ymax=104.5, ytick={80,85,90,95,100}]
\draw[coverageMuted!70, dashed] (axis cs:2000,80) -- (axis cs:2000,104.5);
\addplot[coverageBlue, no marks] coordinates
  {(400,83.16) (1000,96.96) (2000,98.98) (3000,99.32) (4000,99.59)};
\addplot[coverageBlue, smooth, only marks, mark=*, mark options={fill=coverageBlue}]
  coordinates {(400,83.16) (1000,96.96) (2000,98.98) (3000,99.32) (4000,99.59)};
\node[font=\scriptsize, text=coverageBlue!85!black, anchor=west,
  xshift=3pt, fill=white, inner sep=1pt]
  at (axis cs:400,83.16) {83.16};
\node[font=\scriptsize, text=coverageBlue!85!black, anchor=north,
  yshift=-4pt, fill=white, inner sep=1pt]
  at (axis cs:1000,96.96) {96.96};
\node[font=\scriptsize, text=coverageBlue!85!black, anchor=south,
  yshift=2pt, fill=white, inner sep=1pt]
  at (axis cs:2000,98.98) {98.98};
\node[font=\scriptsize, text=coverageBlue!85!black, anchor=north,
  yshift=-4pt, fill=white, inner sep=1pt]
  at (axis cs:3000,99.32) {99.32};
\node[font=\scriptsize, text=coverageBlue!85!black, anchor=south east,
  yshift=2pt, fill=white, inner sep=1pt]
  at (axis cs:4000,99.59) {99.59};
\nextgroupplot[title={(b) Shallow\,$\leftrightarrow$\,Fine-ranking},
  xlabel={Shallow top-$K_1$ ($K_2=400$)},
  ymin=97.2, ymax=98.65, ytick={97.2,97.6,98.0,98.4}]
\draw[coverageMuted!70, dashed] (axis cs:2000,97.2) -- (axis cs:2000,98.65);
\addplot[coverageTeal, no marks] coordinates
  {(400,97.34) (1000,98.13) (2000,98.27) (3000,98.33) (4000,98.36)};
\addplot[coverageTeal, only marks, mark=square*, mark options={fill=coverageTeal}]
  coordinates {(400,97.34) (1000,98.13) (2000,98.27) (3000,98.33) (4000,98.36)};
\node[font=\scriptsize, text=coverageTeal!85!black, anchor=west,
  xshift=3pt, fill=white, inner sep=1pt]
  at (axis cs:400,97.34) {97.34};
\node[font=\scriptsize, text=coverageTeal!85!black, anchor=north,
  yshift=-10pt, fill=white, inner sep=1pt]
  at (axis cs:1000,98.13) {98.13};
\node[font=\scriptsize, text=coverageTeal!85!black, anchor=south,
  yshift=2pt, fill=white, inner sep=1pt]
  at (axis cs:2000,98.27) {98.27};
\node[font=\scriptsize, text=coverageTeal!85!black, anchor=north,
  yshift=-4pt, fill=white, inner sep=1pt]
  at (axis cs:3000,98.33) {98.33};
\node[font=\scriptsize, text=coverageTeal!85!black, anchor=south east,
  yshift=2pt, fill=white, inner sep=1pt]
  at (axis cs:4000,98.36) {98.36};
\nextgroupplot[title={(c) Deep\,$\leftrightarrow$\,Fine-ranking},
  xlabel={Deep top-$K_2$ ($K_1=2$k)},
  xmin=-30, xmax=440, ymin=35, ymax=108,
  xtick={10,100,200,400}, xticklabels={10,100,200,400},
  ytick={40,60,80,100}]
\addplot[coverageAmber, mark=square*, mark options={fill=coverageAmber}]
  coordinates {(10,43.16) (20,60.91) (40,75.95)
    (100,90.19) (200,96.12) (400,98.27)};
\node[text=coverageAmber!90!black, anchor=west, xshift=4pt, yshift=-2pt,
  fill=white, inner sep=1pt, font=\scriptsize] at (axis cs:10,43.16) {43.16};
\node[text=coverageAmber!90!black, anchor=west, xshift=4pt, yshift=-2pt,
  fill=white, inner sep=1pt, font=\scriptsize] at (axis cs:20,60.91) {60.91};
\node[text=coverageAmber!90!black, anchor=west, xshift=4pt, yshift=-2pt,
  fill=white, inner sep=1pt, font=\scriptsize] at (axis cs:40,75.95) {75.95};
\node[text=coverageAmber!90!black, anchor=south, yshift=3pt,
  fill=white, inner sep=1pt, font=\scriptsize] at (axis cs:100,90.19) {90.19};
\node[text=coverageAmber!90!black, anchor=south, yshift=3pt,
  fill=white, inner sep=1pt, font=\scriptsize] at (axis cs:200,96.12) {96.12};
\node[text=coverageAmber!90!black, anchor=north east, yshift=-3pt,
  fill=white, inner sep=1pt, font=\scriptsize] at (axis cs:400,98.27) {98.27};
\nextgroupplot[title={(d) Candidate Budget Trade-off},
  minor xtick={},
  xmin=700, xmax=4300,
  xtick={1000,2000,3000,4000}, xticklabels={1000,2000,3000,4000},
  xlabel={Shallow top-$K_1$ ($K_2=400$)},
  ymin=1.5, ymax=2.6, ytick={1.5,2.0,2.5}, yticklabels={1.5,2.0,2.5},
  ylabel={RPM uplift (\%)},
  axis y line*=left, axis x line*=bottom,
  y axis line style={draw=budgetRPM, line width=0.5pt},
  y tick style={budgetRPM},
  yticklabel style={text=budgetRPM!85!black},
  ylabel style={text=budgetRPM!85!black},
  scaled y ticks=false,
  tick label style={font=\scriptsize}, label style={font=\scriptsize},
  x axis line style={draw=budgetFrame, line width=0.45pt},
  x tick style={draw=budgetFrame},
  xlabel style={text=budgetInk},
  grid=major, grid style={draw=budgetGrid, line width=0.3pt}]
\draw[budgetFrame, line width=0.45pt] (axis cs:700,2.6) -- (axis cs:4300,2.6);
\draw[budgetMuted!70, dashed, line width=0.55pt] (axis cs:2000,1.5) -- (axis cs:2000,2.6);
\addplot[budgetRPM, solid, mark=*, mark size=2pt,
  line width=1pt] coordinates
  {(1000,1.77) (2000,2.14) (3000,2.25) (4000,2.31)};
\node[font=\scriptsize, text=budgetRPM!85!black, anchor=north,
  yshift=-5pt, fill=white, inner sep=1pt]
  at (axis cs:1000,1.77) {1.77};
\node[font=\scriptsize, text=budgetRPM!85!black, anchor=south,
  yshift=5pt, fill=white, inner sep=1pt]
  at (axis cs:2000,2.14) {2.14};
\node[font=\scriptsize, text=budgetRPM!85!black, anchor=south,
  yshift=5pt, fill=white, inner sep=1pt]
  at (axis cs:3000,2.25) {2.25};
\node[font=\scriptsize, text=budgetRPM!85!black, anchor=north east,
  xshift=-3pt, yshift=-5pt, fill=white, inner sep=1pt]
  at (axis cs:4000,2.31) {2.31};
\end{groupplot}
\begin{axis}[
  at={(coverage c4r1.south west)}, anchor=south west,
  scale only axis=true, width=0.185\textwidth, height=0.14210526\textwidth,
  xmin=700, xmax=4300,
  xtick={1000,2000,3000,4000}, xticklabels={1000,2000,3000,4000},
  ymin=40, ymax=52, ytick={40,45,50},
  ylabel={P99 (ms)},
  axis y line*=right, axis x line=none,
  y axis line style={draw=budgetLatency, line width=0.5pt},
  y tick style={budgetLatency},
  yticklabel style={text=budgetLatency!85!black},
  ylabel style={text=budgetLatency!85!black},
  scaled y ticks=false,
  tick label style={font=\scriptsize}, label style={font=\scriptsize}]
\addplot[budgetLatency, dashed, mark=square*, mark size=2pt,
  line width=1pt] coordinates
  {(1000,43.8) (2000,45.1) (3000,47.2) (4000,50.3)};
\node[font=\scriptsize, text=budgetLatency!85!black, anchor=south,
  yshift=5pt, fill=white, inner sep=1pt]
  at (axis cs:1000,43.8) {43.8};
\node[font=\scriptsize, text=budgetLatency!85!black, anchor=north,
  yshift=-5pt, fill=white, inner sep=1pt]
  at (axis cs:2000,45.1) {45.1};
\node[font=\scriptsize, text=budgetLatency!85!black, anchor=north,
  yshift=-5pt, fill=white, inner sep=1pt]
  at (axis cs:3000,47.2) {47.2};
\node[font=\scriptsize, text=budgetLatency!85!black, anchor=south east,
  xshift=-3pt, yshift=2pt, fill=white, inner sep=1pt]
  at (axis cs:4000,50.3) {50.3};
\fill[white, draw=budgetFrame, line width=0.35pt, rounded corners=1pt]
  (axis cs:2950,40.45) rectangle (axis cs:4220,43.35);
\draw[budgetRPM, solid, line width=0.8pt] (axis cs:3060,42.5) -- (axis cs:3440,42.5);
\fill[budgetRPM] (axis cs:3250,42.5) circle (1.3pt);
\node[anchor=west, font=\fontsize{5.8}{6.5}\selectfont, text=budgetRPM!85!black, inner sep=0pt]
  at (axis cs:3520,42.5) {RPM};
\draw[budgetLatency, dashed, line width=0.8pt] (axis cs:3060,41.25) -- (axis cs:3440,41.25);
\fill[budgetLatency] (axis cs:3205,40.95) rectangle (axis cs:3295,41.55);
\node[anchor=west, font=\fontsize{5.8}{6.5}\selectfont, text=budgetLatency!85!black, inner sep=0pt]
  at (axis cs:3520,41.25) {P99};
\end{axis}
\end{tikzpicture}
\par\vspace{-4pt}
\caption{Candidate-selection consistency and online budget trade-off: (a) Shallow top-$K_1$
retention of Deep top-400 from the full replayed pool ($\mathrm{HR}_{\mathrm{S}\rightarrow\mathrm{D}}@K_1$);
(b) impact of Shallow quota $K_1$ on fine-ranking winner retention ($\mathrm{HR}_{\mathrm{two}\rightarrow\mathrm{R}}@(K_1, K_2)$)
with Deep quota fixed at $K_2=400$;
(c) impact of Deep quota $K_2$ on fine-ranking winner retention with Shallow quota fixed at $K_1=2{,}000$;
(d) dedicated online budget sweep showing RPM uplift relative to Production (solid circles, left axis)
and absolute P99 latency (dashed squares, right axis) as $K_1$ varies with $K_2=400$.}
\label{fig:candidate-selection}
\Description{Four panels show candidate selection and the online budget trade-off. Panel (a) shows Shallow retention of Deep top-400 candidates; panel (b) shows two-round fine-ranking winner retention while varying the Shallow budget with the Deep budget fixed at 400; panel (c) shows two-round fine-ranking winner retention while varying the Deep budget with the Shallow budget fixed at 2,000. Panel (d) is a dual-y-axis line plot with a linear first-round budget
axis with four equally spaced values: 1,000, 2,000, 3,000, and 4,000.
Blue-gray circles on the left RPM-uplift scale show 1.77, 2.14, 2.25, and
2.31 percent relative to Production. Terracotta squares on the right P99
latency scale show 43.8, 45.1, 47.2, and 50.3 milliseconds, respectively.
Lines connect the observed points without fitting. A vertical dashed
line marks the deployed budget of 2,000. The second-round budget is 400.}
\end{figure*}

\subsection{Experimental Setup}
\label{subsec:experimental-setup}
\paragraph{Dataset.} Following COPR~\cite{zhao2023copr}, we use fine-ranking
logs from a large-scale e-commerce advertising system. Each log records,
for a request, the candidates that entered the fine-ranking stage, about 400
per request, together with their fine-ranking ECPM order, bids, and
value predictions. The dataset contains more than one billion requests
sampled over 15 days; the first 14 days are used for training and the
final day for evaluation.
To evaluate candidate selection, we replay the full pre-ranking pool
of each evaluation request and score it at both depths. These pools
contain up to 16,000 candidates per request, matching online serving.
The deployed funnel retains $K_1=2{,}000$ after the shallow round and
forwards $K_2=400$ after the deep round to the fine-ranking stage. The
backbone has $N=1$ prefix layer and $M=3$ suffix layers with a uniform token
dimension of $d=128$ (expanding $128\rightarrow256\rightarrow128$ inside the FFN);
Appendix~\ref{subsec:appendix-model-training} lists the remaining
settings.

\paragraph{Compared methods.} \textbf{Production} is the pre-ranking system
deployed before OmniPreRank. It serves the two tasks with separate
models: the LTR signal is predicted by a Transformer-based pre-ranking
model, and the value quantities that form pGMV by a multi-task MMoE
model~\cite{ma2018mmoe} with DIN-based behavior
modeling~\cite{zhou2018din}. \textbf{Shallow-only} and \textbf{Full Deep}
are single-depth models trained separately on the same inputs and data:
Shallow-only trains the $N$-layer prefix with the shallow heads alone,
and Full Deep trains the full $L$-layer backbone with the deep heads
alone; each scores all candidates in a single round. \textbf{OmniPreRank}
jointly trains both depths and serves them in two rounds. In offline evaluation,
all candidates are scored at both depths, so we report Shallow and Deep outputs of
OmniPreRank separately and compare them with Shallow-only and Full Deep.

\paragraph{Offline metrics.} We measure two types of consistency: prediction consistency on the fine-ranking-log candidates and candidate-selection consistency on the replayed pre-ranking benchmark. 
\begingroup
\settowidth{\leftmargini}{\labelitemi}
\addtolength{\leftmargini}{\labelsep}
\begin{list}{\labelitemi}{%
  \setlength{\leftmargin}{\leftmargini}%
  \setlength{\labelwidth}{\leftmargin}%
  \addtolength{\labelwidth}{-\labelsep}%
  \setlength{\topsep}{3pt}%
  \setlength{\partopsep}{0pt}%
  \setlength{\itemsep}{1pt}%
  \setlength{\parsep}{0pt}}
\item \textbf{Prediction consistency.} For LTR, the rankscore is the predicted LTR score multiplied by the logged pre-ranking bid of the candidate.
Within each request, we compute AUC over the fine-ranking-stage candidates using rankscores, with the \emph{fine-ranking winners}, i.e., the head candidates that win the fine-ranking auction and enter re-ranking, as positives and the other candidates as negatives.
We report gAUC as the mean of these request-level AUCs.
For CVR and AMT, we report the MSE loss in Eq.~\eqref{eq:value-loss}.
\item \textbf{Candidate-selection consistency.} On the replayed pre-ranking
benchmark, we measure the consistency between the candidate sets selected
by OmniPreRank and their downstream reference sets by HitRate (HR).
For each request $q$, let $\mathcal{S}_q$ denote the selected set and
$\mathcal{T}_q$ the downstream reference set:
\begin{equation}
\mathrm{HR}=\frac{1}{|\mathcal{Q}|}\sum_{q\in\mathcal{Q}}
\frac{|\mathcal{S}_q\cap\mathcal{T}_q|}{|\mathcal{T}_q|}.
\label{eq:reference-retention}
\end{equation}
$\mathrm{HR}_{\mathrm{S}\rightarrow\mathrm{D}}@K_1$ measures Shallow--Deep
consistency, taking the Shallow top-$K_1$ as $\mathcal{S}_q$ and the Deep
top-400 over all replayed candidates as $\mathcal{T}_q$.
$\mathrm{HR}_{\mathrm{two}\rightarrow\mathrm{R}}@(K_1, K_2)$ measures two-round
cascade consistency with the fine-ranking stage: Shallow first retains $K_1$
candidates, Deep selects the top-$K_2$ set $\mathcal{S}_q$ from that retained
pool, and the reference set $\mathcal{T}_q$ consists of the top-10 fine-ranking
winners recorded in the production logs for request $q$.
\end{list}
\endgroup

\paragraph{Serving metrics.} We measure serving cost through end-to-end
load tests on NVIDIA RTX PRO 5000 Blackwell GPUs (72~GB of memory;
hereafter Pro5000) under the same hardware
and production average batch-size distribution (with packages capped at
1,024 candidates), so the measured P99 latency matches the online
latency of each variant. For each variant, we report maximum sustainable QPS per
GPU and the corresponding serving P99 latency when the load reaches
the same hardware utilization threshold. Here, QPS counts complete package
throughput per second (including both Round~1 and Round~2 for two-round
variants). We also report GPU
memory and inference FLOPs in GFLOPs per request. Bidding latency is
reported separately, as P99 per call in each round.

\paragraph{Online metrics.} Online A/B tests report the relative changes
in revenue per mille (RPM), return on investment (ROI), and click-through rate
(CTR) against Production, together with online P99 latency and the
relative change in prediction fleet size computed under equal total request
load (total QPS divided by per-machine sustainable QPS). All online
variants, including the online budget sweep, are evaluated with 10\% of
traffic per group over 7 days.

\subsection{Offline Consistency}
\label{subsec:prediction-selection-quality}
Table~\ref{tab:offline-consistency} reports prediction consistency with
the fine-ranking stage. OmniPreRank (Shallow)
achieves $2.05$ pp gAUC gain over Production, and
reduces CVR and AMT MSEs by $28.2\%$ and $18.0\%$, respectively.
OmniPreRank (Deep) further improves consistency, reaching gAUC $0.8596$
with CVR and AMT MSE reductions of $33.4\%$ and $24.3\%$ relative to
Production, achieving the highest gAUC and lowest value MSEs among
the compared methods.

We also compare the jointly trained Shallow and Deep outputs with
their separately trained counterparts. OmniPreRank (Shallow) achieves higher gAUC
and lower CVR and AMT MSEs than Shallow-only. With the same backbone
architecture, OmniPreRank (Deep) achieves gAUC $0.8596$ versus $0.8563$
for Full Deep and reduces AMT MSE by $13.8\%$, with nearly unchanged
CVR MSE. These comparisons support the compatibility of shallow and
deep supervision in the shared backbone.

\begin{table}[t]
\centering
\caption{Predictive consistency with the fine-ranking stage.}
\label{tab:offline-consistency}
\small
\setlength{\tabcolsep}{5pt}
\begin{tabular*}{\columnwidth}{@{\extracolsep{\fill}}lccc@{}}
\toprule
Method & gAUC$\uparrow$ & CVR MSE$\downarrow$ & AMT MSE$\downarrow$ \\
\midrule
Production & 0.8289 & 0.57935 & 0.06140 \\
\midrule
Shallow-only & 0.8422 & 0.43827 & 0.05592 \\
OmniPreRank (Shallow) & 0.8494 & 0.41586 & 0.05037 \\
\midrule
Full Deep & 0.8563 & 0.38642 & 0.05391 \\
OmniPreRank (Deep) & \textbf{0.8596} & \textbf{0.38605} & \textbf{0.04645} \\
\bottomrule
\end{tabular*}
\end{table}

Figure~\ref{fig:candidate-selection}(a) evaluates Shallow--Deep
consistency ($\mathrm{HR}_{\mathrm{S}\rightarrow\mathrm{D}}@K_1$) by comparing
Shallow top-$K_1$ with Deep top-400 over all replayed candidates. Retention
rises from $83.16\%$ at $K_1=400$ to $96.96\%$ at $K_1=1{,}000$, $98.98\%$ at
the deployed $K_1=2{,}000$, and $99.59\%$ at $K_1=4{,}000$. Thus, the deployed
Shallow budget preserves approximately $99\%$ of Deep reference candidates.

Figure~\ref{fig:candidate-selection}(b,c) evaluates two-round consistency with
fine-ranking ($\mathrm{HR}_{\mathrm{two}\rightarrow\mathrm{R}}@(K_1, K_2)$) by
varying one budget while fixing the other at its deployed
value. With $K_2=400$, HR rises from $97.34\%$ at $K_1=400$ to
$98.27\%$ at $K_1=2{,}000$ and $98.36\%$ at $K_1=4{,}000$, showing
marginal gains beyond 2,000. With $K_1=2{,}000$, HR against the fixed
top-10 fine-ranking winners reaches $90.19\%$, $96.12\%$, and $98.27\%$
at $K_2=100$, 200, and 400, respectively. Thus, the Deep budget remains
important for winner retention even when Shallow preserves nearly all
Deep reference candidates.

\subsection{Serving Cost Breakdown}
\label{subsec:serving-efficiency}
Table~\ref{tab:serving-breakdown} compares serving performance on
Pro5000 GPUs (72~GB) with identical operator optimizations, measuring
QPS and P99 at the same hardware utilization threshold. OmniPreRank
sustains 1,950 QPS at 45~ms P99, compared with 800 QPS at 56~ms for
Full Deep: 2.44$\times$ the throughput and 11~ms lower P99 latency.
Two-round recomputation reaches 1,200 QPS at 63~ms P99. With the
checkpoint, retained candidates, and precision fixed, state reuse
increases throughput by 62.5\% and lowers P99 by 18~ms
(Table~\ref{tab:serving-breakdown}).

\begin{table}[t]
\centering
\caption{Offline serving cost on Pro5000 GPUs from end-to-end load tests.
QPS measures complete package throughput (packages/s/GPU, including both
rounds for two-round variants) under the production average batch-size
distribution. Production covers only its LTR model.}
\label{tab:serving-breakdown}
\small
\setlength{\tabcolsep}{3.5pt}
\begin{tabular}{@{}lcccc@{}}
\toprule
Execution & QPS & P99 (ms) & \shortstack{GPU memory\\(GB)} &
\shortstack{GFLOPs\\per request} \\
\midrule
Production & 2,100 & 40 & 8 & 300 \\
Full Deep & 800 & 56 & 12 & 1,250 \\
Two rounds, recompute & 1,200 & 63 & 10 & 460 \\
OmniPreRank & 1,950 & 45 & 12.8 & 420 \\
\bottomrule
\end{tabular}
\end{table}

The computation savings follow Section~\ref{subsec:cost-analysis} under the
$16{,}000\rightarrow2{,}000$ funnel (12.5\% retention): Full
Deep requires 1,250 GFLOPs per request, two-round recomputation about 460, and
OmniPreRank about 420 (300 for Shallow over 16,000 candidates and about 120 for
Deep over 2,000 retained candidates), a 66.4\% reduction
obtained mainly by limiting deep computation to retained candidates.
Reuse also avoids repeated prefix computation on the 2,000 retained candidates,
saving about 40 GFLOPs (8.7\%) over recomputation. FLOPs measure arithmetic only; throughput also
depends on feature access, data transfer, and RPC overhead, so FLOP
ratios need not translate into throughput gains. Non-cache GPU memory
reflects loaded execution graphs, model weights, and batch-dependent runtime
activations, which vary across variants (12~GB for Full Deep and 10~GB for
two-round recomputation). At the benchmark load, OmniPreRank's estimated GPU
memory is 12.8~GB: 8~GB for its split execution graphs, weights, and runtime
activations plus up to 4.8~GB for the resident cache. The 15~GB cache hard cap
permits up to 23~GB for serving and cache on the 72~GB Pro5000.

The Production row covers only its LTR model; the separate GMV model for
pGMV needs 70\% as many machines, so the combined baseline is 170\% of
the LTR fleet. Computed under equal total request load (total QPS divided by
per-machine sustainable QPS in Table~\ref{tab:serving-breakdown}), OmniPreRank
unifies both tasks at both depths and uses 36.7\% fewer prediction machines than
this baseline, whereas Full Deep requires 54.4\% more
(Section~\ref{subsec:online-evaluation}). Same-data-center GPU bidding also cuts
P99 from 10 to 2~ms per round (80\%) relative to remote CPU bidding,
supporting two bidding rounds within the latency budget.
A package that misses the deep-round deadline falls back to Shallow scores
(Section~\ref{subsec:cache-capacity-fallback}); in production, such fallback
affects only a small fraction of packages.

\subsection{Online Evaluation}
\label{subsec:online-evaluation}
Table~\ref{tab:online-results} reports online A/B results over 7 days with
10\% of traffic per group, relative to Production (40~ms P99).
OmniPreRank (Shallow) runs the shallow output alone, lowering P99 to
38~ms while gaining RPM $+1.29\%$, ROI $+0.77\%$, and CTR $+0.95\%$.
OmniPreRank (Deep) changes only online inference, running the deep output
alone over all candidates in one round ($K_1=16{,}000$); it yields the
largest gains (RPM $+2.35\%$, ROI $+1.22\%$, CTR $+1.78\%$) but raises P99
to 56~ms and, executing the same computation as Full Deep, needs 54.4\%
more machines than the combined LTR+GMV Production fleet. The full
pipeline reaches RPM $+2.14\%$, ROI $+1.09\%$, and CTR $+1.34\%$ at 45~ms
P99, retaining 91.1\% of the Deep variant's RPM gain with 36.7\% fewer
prediction machines (Section~\ref{subsec:serving-efficiency}).

\begin{table}[t]
\centering
\caption{Online A/B results against Production. $\Delta$Machines is the change in
prediction fleet size relative to Production (LTR+GMV), computed under equal
total request load from per-machine sustainable QPS. All variants ran over
7 days with 10\% traffic per group.}
\label{tab:online-results}
\small
\setlength{\tabcolsep}{2.5pt}
\begin{tabular*}{\columnwidth}{@{\extracolsep{\fill}}lccccc@{}}
\toprule
Method & $\Delta$RPM & $\Delta$ROI & $\Delta$CTR & P99 (ms) &
$\Delta$Machines \\
\midrule
OmniPreRank (Shallow) & $+1.29\%$ & $+0.77\%$ & $+0.95\%$ & 38 & n/a \\
OmniPreRank (Deep) & $+2.35\%$ & $+1.22\%$ & $+1.78\%$ & 56 & $+54.4\%$ \\
\textbf{OmniPreRank} & $+2.14\%$ & $+1.09\%$ & $+1.34\%$ & 45 & $-36.7\%$ \\
\bottomrule
\end{tabular*}
\end{table}

\subsection{Candidate Budget Trade-off}
\label{subsec:budget-tradeoff}
Figure~\ref{fig:candidate-selection}(d) varies the first-round budget $K_1$
in a dedicated online budget sweep (7 days, 10\% traffic per group) with
$K_2=400$. Raising $K_1$ from 1,000 to 2,000 increases the RPM
uplift relative to Production from 1.77\% to 2.14\%, at a P99 latency cost of
1.3~ms. The next two increases of 1,000 candidates add 2.1 and 3.1~ms,
while the incremental RPM gains diminish from 0.37 to 0.11 and 0.06
percentage points. Moving from 2,000 to 4,000 thus adds only 0.17
percentage points of RPM uplift for 5.2~ms of additional P99 latency.
This diminishing online gain is consistent with the offline HR trend in
Figure~\ref{fig:candidate-selection}(a,b), where retention approaches saturation
as the Shallow budget increases. This observed trade-off supports $K_1=2{,}000$
as the selected deployment compromise.

\section{Related Work}
\label{sec:related-work}

\subsection{Toward Unified Ranking}
\label{subsec:unified-recommendation-reuse}
Recent work moves recommendation toward fewer and larger models.
OneRec replaces the retrieve-and-rank cascade with a generative recommender trained with preference alignment~\cite{deng2025onerec}.
Within a ranking stage, OneTrans and EST unify sequence modeling and heterogeneous feature interactions in a single Transformer~\cite{zhang2025onetrans,liu2026est}; our backbone builds on them.
Concurrently with our work, OneTrans-V2 serves retrieval, pre-ranking, and fine-ranking with one Transformer while retaining the cascade~\cite{cao2026onetransv2}.
OmniPreRank pursues unification differently: instead of replacing the cascade or unifying stage-level models, it keeps the decision role of each filter and merges computation across one filter into a single model.

\subsection{Computation Reuse and Progressive Computation}
\label{subsec:computation-reuse}
Several systems reuse computation to make larger models affordable.
HSTU's M-FALCON reuses user-history computation across candidate microbatches within a ranking stage~\cite{zhai2024hstu}.
OneTrans-V2 encodes candidate-independent user context once and shares it across stages, while candidate features and their computation remain stage-specific~\cite{cao2026onetransv2}.
UniSGR reuses semantic identifiers and decoder states produced during generative retrieval in subsequent ranking~\cite{sun2026unisgr}.
In score-based cascades, candidate-dependent computation is typically repeated at successive stages; OmniPreRank instead resumes it from retained candidates' cached states.

Progressive computation within one network has also been studied outside recommendation.
Early-exit models attach classifiers to intermediate layers and stop the forward pass of an input once its prediction is confident or stable~\cite{xin2020deebert,zhou2020pabee}.
Closest in form, the Cascade Transformer attaches rankers at several layers of one Transformer, prunes candidates between them, and continues the remaining candidates from their partial encodings~\cite{soldaini2020cascade}.
In both lines of work, the model itself decides, per input or by its own scores, where computation stops within one forward pass.
In OmniPreRank, the filter lies outside the model: it uses bid-dependent ECPM and business rules applied after engine-side aggregation, so each depth must predict both the ranking signal and pGMV, and the states must persist across a separate request (Section~\ref{sec:online-deployment}).

\subsection{Pre-Ranking in Cascaded Systems}
\label{subsec:cascade-ranking-learning}
Multi-stage systems apply increasingly expensive models to smaller candidate sets~\cite{covington2016youtube,wang2020cold}.
One line of pre-ranking work improves the quality attainable under a per-candidate budget: COLD jointly designs the model and its serving~\cite{wang2020cold}, IntTower strengthens feature interactions in a two-tower architecture~\cite{li2022inttower}, and TARQ approximates target attention with residual quantization~\cite{li2025tarq}.
Another line aligns an early stage with downstream selection: COPR aligns pre-ranking with the fine-ranking ECPM order while accounting for bids~\cite{zhao2023copr}, FS-LTR models downstream selection preferences~\cite{zheng2024fsltr}, and LCRON optimizes the probability that target items survive the whole cascade~\cite{wang2025lcron}.
OmniPreRank adopts COPR's objective and changes the execution structure rather than the training target.
The closest work within pre-ranking is HAP, which scores the initial pool with a lightweight model and the survivors with a more expressive one~\cite{tong2026hap}, but uses separate models; OmniPreRank instead continues the first round's computation within one model.

\section{Conclusion}
\label{sec:conclusion}
OmniPreRank takes a first step toward unified ranking by merging the computation on both sides of a filter inside advertising pre-ranking into one model.
A unified multi-task Transformer predicts the ranking signal and pGMV at a shallow and a deep depth: the shallow round supplies the bids for the first filter and builds the states from which the deep round continues, while same-data-center GPU bidding and GPU state caching with request routing carry this computation across bidding, engine-side filtering, and a separate request.
Offline, both depths are more consistent with fine-ranking than the production model.
Online, OmniPreRank increases RPM by 2.14\%, retaining 91.1\% of the RPM gain of full-candidate deep scoring, while using 36.7\% fewer prediction machines than the complete production fleet serving separate LTR and GMV models under equal request load, and it is fully deployed in production.

The deep round can currently refine only the inputs encoded in the first round, and under fallback, Shallow and Deep scores are ranked together without calibration.

\section{Outlook}
\paragraph{Toward unified ranking.}
The two-round design suggests a gradual transition from separate stage models to successive decisions at different depths of one model.
A natural next step is to extend the deep round into fine-ranking: retained candidates would carry their pre-ranking states through the filter and continue computation, with each depth providing the predictions needed for its bidding and selection decisions.
Features that are too costly to process for the full pool could be introduced as additional tokens after filtering, allowing later layers to combine richer information with the retained states.
This extension requires aligning the stages' features and training objectives, preparing early representations for later refinement, and supporting additional inputs during serving.
Further depths could be added where additional decisions are needed.
Evaluation would need to establish the resulting selection quality and total cost, including feature access and state management, beyond the single filter studied here.

\paragraph{Coordinating model and external computation.}
GPU state caching and GPU bidding suggest a broader opportunity to coordinate model inference with the surrounding decision pipeline.
Prediction, bidding, and candidate selection operate on overlapping candidate data and produce inputs for one another.
Moving suitable external operations onto GPUs could let them exchange results directly and reuse data resident in GPU memory across decision points.
A runtime organized around these data dependencies could coordinate model graphs, engine operations, communication, and state reclamation, reducing repeated preparation and transfer of candidate data.
Such coordination would need to accommodate aggregation across serving shards, changing candidate sets, and irregular business logic, while sharing GPU memory and compute resources with inference.
Its value should be assessed through the latency and resource cost of the full pipeline under the same decision policies and reliability requirements.
OmniPreRank provides an initial example through GPU bidding and state reuse across requests; extending this coordination to additional engine operations remains future work.

\bibliographystyle{ACM-Reference-Format}
\setlength{\bibsep}{0pt plus 1pt}
\bibliography{references}


\begin{thebibliography}{32}


\ifx \showCODEN    \undefined \def \showCODEN     #1{\unskip}     \fi
\ifx \showISBNx    \undefined \def \showISBNx     #1{\unskip}     \fi
\ifx \showISBNxiii \undefined \def \showISBNxiii  #1{\unskip}     \fi
\ifx \showISSN     \undefined \def \showISSN      #1{\unskip}     \fi
\ifx \showLCCN     \undefined \def \showLCCN      #1{\unskip}     \fi
\ifx \shownote     \undefined \def \shownote      #1{#1}          \fi
\ifx \showarticletitle \undefined \def \showarticletitle #1{#1}   \fi
\ifx \showURL      \undefined \def \showURL       {\relax}        \fi
\providecommand\bibfield[2]{#2}
\providecommand\bibinfo[2]{#2}
\providecommand\natexlab[1]{#1}
\providecommand\showeprint[2][]{arXiv:#2}
\makeatletter
\@ifundefined{NAT@parse@date}{}{\let\NAT@parse@date@orig\NAT@parse@date}
\@ifundefined{NAT@parse@date}{}{\def\NAT@parse@date#1#2#3#4#5#6@@{\NAT@parse@date@orig#1#2#3#4#5#6@@\def\NAT@tempyear{0000}\def\NAT@tempexlab{{?}}\ifx\NAT@year\NAT@tempyear\ifx\NAT@exlab\NAT@tempexlab\def\NAT@date{[n.\,d.]}\else\edef\NAT@date{[n.\,d.]\NAT@exlab}\fi\fi}}
\makeatother

\bibitem[Bian et~al\mbox{.}(2022)]%
        {bian2022can}
\bibfield{author}{\bibinfo{person}{Weijie Bian}, \bibinfo{person}{Kailun Wu},
  \bibinfo{person}{Lejian Ren}, \bibinfo{person}{Qi Pi},
  \bibinfo{person}{Yujing Zhang}, \bibinfo{person}{Can Xiao},
  \bibinfo{person}{Xiang-Rong Sheng}, \bibinfo{person}{Yong-Nan Zhu},
  \bibinfo{person}{Zhangming Chan}, \bibinfo{person}{Na Mou},
  \bibinfo{person}{Xinchen Luo}, \bibinfo{person}{Shiming Xiang},
  \bibinfo{person}{Guorui Zhou}, \bibinfo{person}{Xiaoqiang Zhu}, {and}
  \bibinfo{person}{Hongbo Deng}.} \bibinfo{year}{2022}\natexlab{}.
\newblock \showarticletitle{{CAN}: Feature Co-Action Network for Click-Through
  Rate Prediction}. In \bibinfo{booktitle}{\emph{Proceedings of the 15th ACM
  International Conference on Web Search and Data Mining}}.
  \bibinfo{publisher}{Association for Computing Machinery},
  \bibinfo{address}{New York, NY, USA}, \bibinfo{pages}{57--65}.
\newblock


\bibitem[Birrell and Nelson(1984)]%
        {birrell1984rpc}
\bibfield{author}{\bibinfo{person}{Andrew~D. Birrell} {and}
  \bibinfo{person}{Bruce~Jay Nelson}.} \bibinfo{year}{1984}\natexlab{}.
\newblock \showarticletitle{Implementing Remote Procedure Calls}.
\newblock \bibinfo{journal}{\emph{ACM Transactions on Computer Systems}}
  \bibinfo{volume}{2}, \bibinfo{number}{1} (\bibinfo{year}{1984}),
  \bibinfo{pages}{39--59}.
\newblock
\href{https://doi.org/10.1145/2080.357392}{doi:\nolinkurl{10.1145/2080.357392}}


\bibitem[Cao et~al\mbox{.}(2026)]%
        {cao2026onetransv2}
\bibfield{author}{\bibinfo{person}{Hannan Cao}, \bibinfo{person}{Jun Guo},
  \bibinfo{person}{Haolei Pei}, \bibinfo{person}{Zhaoqi Zhang},
  \bibinfo{person}{Tianyu Wang}, \bibinfo{person}{Ziyang Wang},
  \bibinfo{person}{Youchen Sun}, \bibinfo{person}{Yue Xue},
  \bibinfo{person}{Yucheng Mao}, \bibinfo{person}{Lintao Yan},
  \bibinfo{person}{Yufei Feng}, \bibinfo{person}{Shaowei Liu},
  \bibinfo{person}{Rongkun Xing}, \bibinfo{person}{Feiling Gong},
  \bibinfo{person}{Xinyu Chenli}, \bibinfo{person}{Cong Xu},
  \bibinfo{person}{Mingge Zhang}, \bibinfo{person}{Yunjia Zhu},
  \bibinfo{person}{Yajing Zhang}, \bibinfo{person}{Pengfei Ren}, {and}
  \bibinfo{person}{Yue Lin}.} \bibinfo{year}{2026}\natexlab{}.
\newblock \bibinfo{title}{{OneTrans-V2}: Unifying Retrieval, Pre-rank, and
  Fine-rank with One Transformer in Industrial Recommender}.
\newblock
\showeprint[arxiv]{2609.28589}~[cs.IR]
\urldef\tempurl%
\url{https://arxiv.org/abs/2609.28589}
\showURL{%
\tempurl}


\bibitem[Chan et~al\mbox{.}(2025)]%
        {chan2025mal}
\bibfield{author}{\bibinfo{person}{Zhangming Chan}, \bibinfo{person}{Sishuo
  Chen}, \bibinfo{person}{Xiang-Rong Sheng}, \bibinfo{person}{Lei Zhang},
  \bibinfo{person}{Sheng Chen}, \bibinfo{person}{Chenghuan Hou},
  \bibinfo{person}{Han Zhu}, \bibinfo{person}{Jian Xu}, {and}
  \bibinfo{person}{Bo Zheng}.} \bibinfo{year}{2025}\natexlab{}.
\newblock \showarticletitle{See Beyond a Single View: Multi-Attribution
  Learning Leads to Better Conversion Rate Prediction}. In
  \bibinfo{booktitle}{\emph{Proceedings of the 34th ACM International
  Conference on Information and Knowledge Management}}.
  \bibinfo{publisher}{Association for Computing Machinery},
  \bibinfo{address}{New York, NY, USA}, \bibinfo{pages}{5600--5608}.
\newblock


\bibitem[Chan et~al\mbox{.}(2023a)]%
        {chan2023pssa}
\bibfield{author}{\bibinfo{person}{Zhangming Chan}, \bibinfo{person}{Jiacen
  Hu}, \bibinfo{person}{Yu Zhang}, \bibinfo{person}{Shuguang Han},
  \bibinfo{person}{Siyuan Lou}, \bibinfo{person}{Baolin Liu},
  \bibinfo{person}{Han Zhu}, \bibinfo{person}{Yuning Jiang},
  \bibinfo{person}{Jian Xu}, {and} \bibinfo{person}{Bo Zheng}.}
  \bibinfo{year}{2023}\natexlab{a}.
\newblock \showarticletitle{{PS-SA}: An Efficient Self-Attention via
  Progressive Sampling for User Behavior Sequence Modeling}. In
  \bibinfo{booktitle}{\emph{Proceedings of the 32nd ACM International
  Conference on Information and Knowledge Management}}.
  \bibinfo{publisher}{Association for Computing Machinery},
  \bibinfo{address}{New York, NY, USA}, \bibinfo{pages}{4639--4645}.
\newblock


\bibitem[Chan et~al\mbox{.}(2023b)]%
        {chan2023hdr}
\bibfield{author}{\bibinfo{person}{Zhangming Chan}, \bibinfo{person}{Yu Zhang},
  \bibinfo{person}{Shuguang Han}, \bibinfo{person}{Yong Bai},
  \bibinfo{person}{Xiang-Rong Sheng}, \bibinfo{person}{Siyuan Lou},
  \bibinfo{person}{Jiacen Hu}, \bibinfo{person}{Baolin Liu},
  \bibinfo{person}{Yuning Jiang}, \bibinfo{person}{Jian Xu}, {and}
  \bibinfo{person}{Bo Zheng}.} \bibinfo{year}{2023}\natexlab{b}.
\newblock \showarticletitle{Capturing Conversion Rate Fluctuation during Sales
  Promotions: A Novel Historical Data Reuse Approach}. In
  \bibinfo{booktitle}{\emph{Proceedings of the 29th ACM SIGKDD Conference on
  Knowledge Discovery and Data Mining}}. \bibinfo{publisher}{Association for
  Computing Machinery}, \bibinfo{address}{New York, NY, USA},
  \bibinfo{pages}{3774--3784}.
\newblock


\bibitem[Chan et~al\mbox{.}(2022)]%
        {chan2022keep}
\bibfield{author}{\bibinfo{person}{Zhangming Chan}, \bibinfo{person}{Yujing
  Zhang}, \bibinfo{person}{Shuhao Xu}, \bibinfo{person}{Weijie Bian},
  \bibinfo{person}{Shuguang Han}, \bibinfo{person}{Hongbo Deng}, {and}
  \bibinfo{person}{Bo Zheng}.} \bibinfo{year}{2022}\natexlab{}.
\newblock \showarticletitle{{KEEP}: An Industrial Pre-Training Framework for
  Online Recommendation via Knowledge Extraction and Plugging}. In
  \bibinfo{booktitle}{\emph{Proceedings of the 31st ACM International
  Conference on Information and Knowledge Management}}.
  \bibinfo{publisher}{Association for Computing Machinery},
  \bibinfo{address}{New York, NY, USA}, \bibinfo{pages}{3684--3693}.
\newblock


\bibitem[Covington et~al\mbox{.}(2016)]%
        {covington2016youtube}
\bibfield{author}{\bibinfo{person}{Paul Covington}, \bibinfo{person}{Jay
  Adams}, {and} \bibinfo{person}{Emre Sargin}.}
  \bibinfo{year}{2016}\natexlab{}.
\newblock \showarticletitle{Deep Neural Networks for {YouTube}
  Recommendations}. In \bibinfo{booktitle}{\emph{Proceedings of the 10th ACM
  Conference on Recommender Systems}}. \bibinfo{publisher}{Association for
  Computing Machinery}, \bibinfo{address}{New York, NY, USA},
  \bibinfo{pages}{191--198}.
\newblock
\href{https://doi.org/10.1145/2959100.2959190}{doi:\nolinkurl{10.1145/2959100.2959190}}


\bibitem[Deng et~al\mbox{.}(2025)]%
        {deng2025onerec}
\bibfield{author}{\bibinfo{person}{Jiaxin Deng}, \bibinfo{person}{Shiyao Wang},
  \bibinfo{person}{Kuo Cai}, \bibinfo{person}{Lejian Ren},
  \bibinfo{person}{Qigen Hu}, \bibinfo{person}{Weifeng Ding},
  \bibinfo{person}{Qiang Luo}, {and} \bibinfo{person}{Guorui Zhou}.}
  \bibinfo{year}{2025}\natexlab{}.
\newblock \bibinfo{title}{{OneRec}: Unifying Retrieve and Rank with Generative
  Recommender and Iterative Preference Alignment}.
\newblock
\showeprint[arxiv]{2502.18965}~[cs.IR]
\urldef\tempurl%
\url{https://arxiv.org/abs/2502.18965}
\showURL{%
\tempurl}


\bibitem[Jin et~al\mbox{.}(2026)]%
        {jin2026tga}
\bibfield{author}{\bibinfo{person}{Hanqi Jin}, \bibinfo{person}{Gaoming Yang},
  \bibinfo{person}{Zhangming Chan}, \bibinfo{person}{Yapeng Yuan},
  \bibinfo{person}{Longbin Li}, \bibinfo{person}{Fei Sun},
  \bibinfo{person}{Yeqiu Yang}, \bibinfo{person}{Jian Wu},
  \bibinfo{person}{Yuning Jiang}, {and} \bibinfo{person}{Bo Zheng}.}
  \bibinfo{year}{2026}\natexlab{}.
\newblock \showarticletitle{Multi-Behavior Sequential Modeling with
  Transition-Aware Graph Attention Network for E-Commerce Recommendation}. In
  \bibinfo{booktitle}{\emph{Proceedings of the ACM Web Conference 2026}}.
  \bibinfo{publisher}{Association for Computing Machinery},
  \bibinfo{address}{New York, NY, USA}, \bibinfo{pages}{8681--8684}.
\newblock
\href{https://doi.org/10.1145/3774904.3792939}{doi:\nolinkurl{10.1145/3774904.3792939}}


\bibitem[Li et~al\mbox{.}(2022)]%
        {li2022inttower}
\bibfield{author}{\bibinfo{person}{Xiangyang Li}, \bibinfo{person}{Bo Chen},
  \bibinfo{person}{HuiFeng Guo}, \bibinfo{person}{Jingjie Li},
  \bibinfo{person}{Chenxu Zhu}, \bibinfo{person}{Xiang Long},
  \bibinfo{person}{Sujian Li}, \bibinfo{person}{Yichao Wang},
  \bibinfo{person}{Wei Guo}, \bibinfo{person}{Longxia Mao},
  \bibinfo{person}{Jinxing Liu}, \bibinfo{person}{Zhenhua Dong}, {and}
  \bibinfo{person}{Ruiming Tang}.} \bibinfo{year}{2022}\natexlab{}.
\newblock \bibinfo{title}{{IntTower}: the Next Generation of Two-Tower Model
  for Pre-Ranking System}.
\newblock
\showeprint[arxiv]{2210.09890}~[cs.IR]
\urldef\tempurl%
\url{https://arxiv.org/abs/2210.09890}
\showURL{%
\tempurl}


\bibitem[Li et~al\mbox{.}(2026)]%
        {li2026gmvdelay}
\bibfield{author}{\bibinfo{person}{Xinyu Li}, \bibinfo{person}{Sishuo Chen},
  \bibinfo{person}{Guipeng Xv}, \bibinfo{person}{Li Zhang},
  \bibinfo{person}{Mingxuan Luo}, \bibinfo{person}{Zhangming Chan},
  \bibinfo{person}{Xiang-Rong Sheng}, \bibinfo{person}{Han Zhu},
  \bibinfo{person}{Jian Xu}, {and} \bibinfo{person}{Chen Lin}.}
  \bibinfo{year}{2026}\natexlab{}.
\newblock \showarticletitle{Delayed Feedback Modeling for Post-Click Gross
  Merchandise Volume Prediction: Benchmark, Insights and Approaches}. In
  \bibinfo{booktitle}{\emph{Proceedings of the ACM Web Conference 2026}}.
  \bibinfo{publisher}{Association for Computing Machinery},
  \bibinfo{address}{New York, NY, USA}, \bibinfo{pages}{6299--6308}.
\newblock


\bibitem[Li et~al\mbox{.}(2025)]%
        {li2025tarq}
\bibfield{author}{\bibinfo{person}{Yutong Li}, \bibinfo{person}{Yu Zhu},
  \bibinfo{person}{Yichen Qiao}, \bibinfo{person}{Ziyu Guan},
  \bibinfo{person}{Lv Shao}, \bibinfo{person}{Tong Liu}, {and}
  \bibinfo{person}{Bo Zheng}.} \bibinfo{year}{2025}\natexlab{}.
\newblock \bibinfo{title}{Equip Pre-ranking with Target Attention by Residual
  Quantization}.
\newblock
\showeprint[arxiv]{2509.16931}~[cs.IR]
\urldef\tempurl%
\url{https://arxiv.org/abs/2509.16931}
\showURL{%
\tempurl}


\bibitem[Liu et~al\mbox{.}(2026)]%
        {liu2026est}
\bibfield{author}{\bibinfo{person}{Mingyang Liu}, \bibinfo{person}{Yong Bai},
  \bibinfo{person}{Zhangming Chan}, \bibinfo{person}{Sishuo Chen},
  \bibinfo{person}{Xiang-Rong Sheng}, \bibinfo{person}{Han Zhu},
  \bibinfo{person}{Jian Xu}, {and} \bibinfo{person}{Xinyang Chen}.}
  \bibinfo{year}{2026}\natexlab{}.
\newblock \showarticletitle{{EST}: Towards Efficient Scaling Laws in
  Click-Through Rate Prediction via Unified Modeling}. In
  \bibinfo{booktitle}{\emph{Proceedings of the 32nd ACM SIGKDD Conference on
  Knowledge Discovery and Data Mining V.2}}. \bibinfo{publisher}{Association
  for Computing Machinery}, \bibinfo{address}{New York, NY, USA},
  \bibinfo{pages}{7656--7666}.
\newblock
\href{https://doi.org/10.1145/3770855.3818400}{doi:\nolinkurl{10.1145/3770855.3818400}}


\bibitem[Luo et~al\mbox{.}(2026)]%
        {luo2026netcvr}
\bibfield{author}{\bibinfo{person}{Mingxuan Luo}, \bibinfo{person}{Guipeng Xv},
  \bibinfo{person}{Sishuo Chen}, \bibinfo{person}{Xinyu Li},
  \bibinfo{person}{Li Zhang}, \bibinfo{person}{Zhangming Chan},
  \bibinfo{person}{Xiang-Rong Sheng}, \bibinfo{person}{Han Zhu},
  \bibinfo{person}{Jian Xu}, \bibinfo{person}{Bo Zheng}, {and}
  \bibinfo{person}{Chen Lin}.} \bibinfo{year}{2026}\natexlab{}.
\newblock \showarticletitle{Modeling Cascaded Delay Feedback for Online Net
  Conversion Rate Prediction: Benchmark, Insights and Solutions}. In
  \bibinfo{booktitle}{\emph{Proceedings of the ACM Web Conference 2026}}.
  \bibinfo{publisher}{Association for Computing Machinery},
  \bibinfo{address}{New York, NY, USA}, \bibinfo{pages}{6252--6263}.
\newblock
\href{https://doi.org/10.1145/3774904.3792327}{doi:\nolinkurl{10.1145/3774904.3792327}}


\bibitem[Ma et~al\mbox{.}(2018)]%
        {ma2018mmoe}
\bibfield{author}{\bibinfo{person}{Jiaqi Ma}, \bibinfo{person}{Zhe Zhao},
  \bibinfo{person}{Xinyang Yi}, \bibinfo{person}{Jilin Chen},
  \bibinfo{person}{Lichan Hong}, {and} \bibinfo{person}{Ed~H. Chi}.}
  \bibinfo{year}{2018}\natexlab{}.
\newblock \showarticletitle{Modeling Task Relationships in Multi-task Learning
  with Multi-gate Mixture-of-Experts}. In \bibinfo{booktitle}{\emph{Proceedings
  of the 24th ACM SIGKDD International Conference on Knowledge Discovery \&
  Data Mining}}. \bibinfo{publisher}{Association for Computing Machinery},
  \bibinfo{address}{New York, NY, USA}, \bibinfo{pages}{1930--1939}.
\newblock
\href{https://doi.org/10.1145/3219819.3220007}{doi:\nolinkurl{10.1145/3219819.3220007}}


\bibitem[Sheng et~al\mbox{.}(2024)]%
        {sheng2024multimodal}
\bibfield{author}{\bibinfo{person}{Xiang-Rong Sheng}, \bibinfo{person}{Feifan
  Yang}, \bibinfo{person}{Litong Gong}, \bibinfo{person}{Biao Wang},
  \bibinfo{person}{Zhangming Chan}, \bibinfo{person}{Yujing Zhang},
  \bibinfo{person}{Yueyao Cheng}, \bibinfo{person}{Yong-Nan Zhu},
  \bibinfo{person}{Tiezheng Ge}, \bibinfo{person}{Han Zhu},
  \bibinfo{person}{Yuning Jiang}, \bibinfo{person}{Jian Xu}, {and}
  \bibinfo{person}{Bo Zheng}.} \bibinfo{year}{2024}\natexlab{}.
\newblock \showarticletitle{Enhancing {Taobao} Display Advertising with
  Multimodal Representations: Challenges, Approaches and Insights}. In
  \bibinfo{booktitle}{\emph{Proceedings of the 33rd ACM International
  Conference on Information and Knowledge Management}}.
  \bibinfo{publisher}{Association for Computing Machinery},
  \bibinfo{address}{New York, NY, USA}, \bibinfo{pages}{4858--4865}.
\newblock


\bibitem[Soldaini and Moschitti(2020)]%
        {soldaini2020cascade}
\bibfield{author}{\bibinfo{person}{Luca Soldaini} {and}
  \bibinfo{person}{Alessandro Moschitti}.} \bibinfo{year}{2020}\natexlab{}.
\newblock \showarticletitle{The {Cascade Transformer}: an Application for
  Efficient Answer Sentence Selection}. In
  \bibinfo{booktitle}{\emph{Proceedings of the 58th Annual Meeting of the
  Association for Computational Linguistics}}. \bibinfo{publisher}{Association
  for Computational Linguistics}, \bibinfo{address}{Online},
  \bibinfo{pages}{5697--5708}.
\newblock
\href{https://doi.org/10.18653/v1/2020.acl-main.504}{doi:\nolinkurl{10.18653/v1/2020.acl-main.504}}


\bibitem[Sun et~al\mbox{.}(2026)]%
        {sun2026unisgr}
\bibfield{author}{\bibinfo{person}{Jiawei Sun}, \bibinfo{person}{Jun Yang},
  \bibinfo{person}{Ziyue Guo}, \bibinfo{person}{Dongyue Xu},
  \bibinfo{person}{Jianan Yan}, \bibinfo{person}{Lifang Deng}, {and}
  \bibinfo{person}{Xiaoyi Zeng}.} \bibinfo{year}{2026}\natexlab{}.
\newblock \bibinfo{title}{{UniSGR}: Unified Framework for Semantic {ID}
  Generation and Ranking}.
\newblock
\showeprint[arxiv]{2607.04068}~[cs.IR]
\urldef\tempurl%
\url{https://arxiv.org/abs/2607.04068}
\showURL{%
\tempurl}


\bibitem[Tong et~al\mbox{.}(2026)]%
        {tong2026hap}
\bibfield{author}{\bibinfo{person}{Pengfei Tong}, \bibinfo{person}{Siyuan
  Chen}, \bibinfo{person}{Chenwei Zhang}, \bibinfo{person}{Bo Wang},
  \bibinfo{person}{Qi Pi}, \bibinfo{person}{Pixun Li}, {and}
  \bibinfo{person}{Zuotao Liu}.} \bibinfo{year}{2026}\natexlab{}.
\newblock \bibinfo{title}{Not All Candidates are Created Equal: A
  Heterogeneity-Aware Approach to Pre-ranking in Recommender Systems}.
\newblock
\showeprint[arxiv]{2603.03770}~[cs.IR]
\urldef\tempurl%
\url{https://arxiv.org/abs/2603.03770}
\showURL{%
\tempurl}


\bibitem[Wang et~al\mbox{.}(2025)]%
        {wang2025lcron}
\bibfield{author}{\bibinfo{person}{Yunli Wang}, \bibinfo{person}{Zhen Zhang},
  \bibinfo{person}{Zhiqiang Wang}, \bibinfo{person}{Zixuan Yang},
  \bibinfo{person}{Yu Li}, \bibinfo{person}{Jian Yang},
  \bibinfo{person}{Shiyang Wen}, \bibinfo{person}{Peng Jiang}, {and}
  \bibinfo{person}{Kun Gai}.} \bibinfo{year}{2025}\natexlab{}.
\newblock \showarticletitle{Learning Cascade Ranking as One Network}. In
  \bibinfo{booktitle}{\emph{Proceedings of the 42nd International Conference on
  Machine Learning}} \emph{(\bibinfo{series}{Proceedings of Machine Learning
  Research}, Vol.~\bibinfo{volume}{267})}. \bibinfo{publisher}{PMLR},
  \bibinfo{pages}{65828--65844}.
\newblock
\urldef\tempurl%
\url{https://proceedings.mlr.press/v267/wang25fc.html}
\showURL{%
\tempurl}


\bibitem[Wang et~al\mbox{.}(2020)]%
        {wang2020cold}
\bibfield{author}{\bibinfo{person}{Zhe Wang}, \bibinfo{person}{Liqin Zhao},
  \bibinfo{person}{Biye Jiang}, \bibinfo{person}{Guorui Zhou},
  \bibinfo{person}{Xiaoqiang Zhu}, {and} \bibinfo{person}{Kun Gai}.}
  \bibinfo{year}{2020}\natexlab{}.
\newblock \bibinfo{title}{{COLD}: Towards the Next Generation of Pre-Ranking
  System}.
\newblock
\showeprint[arxiv]{2007.16122}~[cs.IR]
\urldef\tempurl%
\url{https://arxiv.org/abs/2007.16122}
\showURL{%
\tempurl}


\bibitem[Wu et~al\mbox{.}(2026)]%
        {wu2026muse}
\bibfield{author}{\bibinfo{person}{Bin Wu}, \bibinfo{person}{Feifan Yang},
  \bibinfo{person}{Zhangming Chan}, \bibinfo{person}{Yu-Ran Gu},
  \bibinfo{person}{Jiawei Feng}, \bibinfo{person}{Chao Yi},
  \bibinfo{person}{Xiang-Rong Sheng}, \bibinfo{person}{Han Zhu},
  \bibinfo{person}{Jian Xu}, \bibinfo{person}{Mang Ye}, {and}
  \bibinfo{person}{Bo Zheng}.} \bibinfo{year}{2026}\natexlab{}.
\newblock \showarticletitle{{MUSE}: A Simple Yet Effective Multimodal
  Search-Based Framework for Lifelong User Interest Modeling}. In
  \bibinfo{booktitle}{\emph{Proceedings of the 35th ACM International
  Conference on Information and Knowledge Management}}.
  \bibinfo{publisher}{Association for Computing Machinery},
  \bibinfo{address}{New York, NY, USA}.
\newblock


\bibitem[Xin et~al\mbox{.}(2020)]%
        {xin2020deebert}
\bibfield{author}{\bibinfo{person}{Ji Xin}, \bibinfo{person}{Raphael Tang},
  \bibinfo{person}{Jaejun Lee}, \bibinfo{person}{Yaoliang Yu}, {and}
  \bibinfo{person}{Jimmy Lin}.} \bibinfo{year}{2020}\natexlab{}.
\newblock \showarticletitle{{DeeBERT}: Dynamic Early Exiting for Accelerating
  {BERT} Inference}. In \bibinfo{booktitle}{\emph{Proceedings of the 58th
  Annual Meeting of the Association for Computational Linguistics}}.
  \bibinfo{publisher}{Association for Computational Linguistics},
  \bibinfo{address}{Online}, \bibinfo{pages}{2246--2251}.
\newblock
\href{https://doi.org/10.18653/v1/2020.acl-main.204}{doi:\nolinkurl{10.18653/v1/2020.acl-main.204}}


\bibitem[Yi et~al\mbox{.}(2026)]%
        {yi2026native}
\bibfield{author}{\bibinfo{person}{Chao Yi}, \bibinfo{person}{Feifan Yang},
  \bibinfo{person}{Jiawei Feng}, \bibinfo{person}{Sishuo Chen},
  \bibinfo{person}{Zhangming Chan}, \bibinfo{person}{Xiang-Rong Sheng}, {and}
  \bibinfo{person}{Han Zhu}.} \bibinfo{year}{2026}\natexlab{}.
\newblock \showarticletitle{Native Multimodal Representation Learning for
  Click-Through Rate Prediction in E-Commerce Scenarios}. In
  \bibinfo{booktitle}{\emph{Proceedings of the 35th ACM International
  Conference on Information and Knowledge Management}}.
  \bibinfo{publisher}{Association for Computing Machinery},
  \bibinfo{address}{New York, NY, USA}, \bibinfo{pages}{8704--8711}.
\newblock
\href{https://doi.org/10.1145/3799682.3840096}{doi:\nolinkurl{10.1145/3799682.3840096}}


\bibitem[Zhai et~al\mbox{.}(2024)]%
        {zhai2024hstu}
\bibfield{author}{\bibinfo{person}{Jiaqi Zhai}, \bibinfo{person}{Lucy Liao},
  \bibinfo{person}{Xing Liu}, \bibinfo{person}{Yueming Wang},
  \bibinfo{person}{Rui Li}, \bibinfo{person}{Xuan Cao}, \bibinfo{person}{Leon
  Gao}, \bibinfo{person}{Zhaojie Gong}, \bibinfo{person}{Fangda Gu},
  \bibinfo{person}{Jiayuan He}, \bibinfo{person}{Yinghai Lu}, {and}
  \bibinfo{person}{Yu Shi}.} \bibinfo{year}{2024}\natexlab{}.
\newblock \showarticletitle{Actions Speak Louder than Words: Trillion-Parameter
  Sequential Transducers for Generative Recommendations}. In
  \bibinfo{booktitle}{\emph{Proceedings of the 41st International Conference on
  Machine Learning}} \emph{(\bibinfo{series}{Proceedings of Machine Learning
  Research}, Vol.~\bibinfo{volume}{235})}. \bibinfo{publisher}{PMLR},
  \bibinfo{pages}{58484--58509}.
\newblock
\urldef\tempurl%
\url{https://proceedings.mlr.press/v235/zhai24a.html}
\showURL{%
\tempurl}


\bibitem[Zhang et~al\mbox{.}(2025)]%
        {zhang2025onetrans}
\bibfield{author}{\bibinfo{person}{Zhaoqi Zhang}, \bibinfo{person}{Haolei Pei},
  \bibinfo{person}{Jun Guo}, \bibinfo{person}{Tianyu Wang},
  \bibinfo{person}{Yufei Feng}, \bibinfo{person}{Hui Sun},
  \bibinfo{person}{Shaowei Liu}, {and} \bibinfo{person}{Aixin Sun}.}
  \bibinfo{year}{2025}\natexlab{}.
\newblock \bibinfo{title}{{OneTrans}: Unified Feature Interaction and Sequence
  Modeling with One Transformer in Industrial Recommender}.
\newblock
\showeprint[arxiv]{2510.26104}~[cs.IR]
\urldef\tempurl%
\url{https://arxiv.org/abs/2510.26104}
\showURL{%
\tempurl}


\bibitem[Zhao et~al\mbox{.}(2023)]%
        {zhao2023copr}
\bibfield{author}{\bibinfo{person}{Zhishan Zhao}, \bibinfo{person}{Jingyue
  Gao}, \bibinfo{person}{Yu Zhang}, \bibinfo{person}{Shuguang Han},
  \bibinfo{person}{Siyuan Lou}, \bibinfo{person}{Xiang-Rong Sheng},
  \bibinfo{person}{Zhe Wang}, \bibinfo{person}{Han Zhu},
  \bibinfo{person}{Yuning Jiang}, \bibinfo{person}{Jian Xu}, {and}
  \bibinfo{person}{Bo Zheng}.} \bibinfo{year}{2023}\natexlab{}.
\newblock \showarticletitle{{COPR}: Consistency-Oriented Pre-Ranking for Online
  Advertising}. In \bibinfo{booktitle}{\emph{Proceedings of the 32nd ACM
  International Conference on Information and Knowledge Management}}.
  \bibinfo{publisher}{Association for Computing Machinery},
  \bibinfo{address}{New York, NY, USA}, \bibinfo{pages}{4974--4980}.
\newblock
\href{https://doi.org/10.1145/3583780.3615465}{doi:\nolinkurl{10.1145/3583780.3615465}}


\bibitem[Zheng et~al\mbox{.}(2024)]%
        {zheng2024fsltr}
\bibfield{author}{\bibinfo{person}{Kai Zheng}, \bibinfo{person}{Haijun Zhao},
  \bibinfo{person}{Rui Huang}, \bibinfo{person}{Beichuan Zhang},
  \bibinfo{person}{Na Mou}, \bibinfo{person}{Yanan Niu}, \bibinfo{person}{Yang
  Song}, \bibinfo{person}{Hongning Wang}, {and} \bibinfo{person}{Kun Gai}.}
  \bibinfo{year}{2024}\natexlab{}.
\newblock \bibinfo{title}{Full Stage Learning to Rank: A Unified Framework for
  Multi-Stage Systems}.
\newblock
\showeprint[arxiv]{2405.04844}~[cs.IR]
\urldef\tempurl%
\url{https://arxiv.org/abs/2405.04844}
\showURL{%
\tempurl}


\bibitem[Zhou et~al\mbox{.}(2018)]%
        {zhou2018din}
\bibfield{author}{\bibinfo{person}{Guorui Zhou}, \bibinfo{person}{Xiaoqiang
  Zhu}, \bibinfo{person}{Chenru Song}, \bibinfo{person}{Ying Fan},
  \bibinfo{person}{Han Zhu}, \bibinfo{person}{Xiao Ma},
  \bibinfo{person}{Yanghui Yan}, \bibinfo{person}{Junqi Jin},
  \bibinfo{person}{Han Li}, {and} \bibinfo{person}{Kun Gai}.}
  \bibinfo{year}{2018}\natexlab{}.
\newblock \showarticletitle{Deep Interest Network for Click-Through Rate
  Prediction}. In \bibinfo{booktitle}{\emph{Proceedings of the 24th ACM SIGKDD
  International Conference on Knowledge Discovery \& Data Mining}}.
  \bibinfo{publisher}{Association for Computing Machinery},
  \bibinfo{address}{New York, NY, USA}, \bibinfo{pages}{1059--1068}.
\newblock
\href{https://doi.org/10.1145/3219819.3219823}{doi:\nolinkurl{10.1145/3219819.3219823}}


\bibitem[Zhou et~al\mbox{.}(2020)]%
        {zhou2020pabee}
\bibfield{author}{\bibinfo{person}{Wangchunshu Zhou}, \bibinfo{person}{Canwen
  Xu}, \bibinfo{person}{Tao Ge}, \bibinfo{person}{Julian McAuley},
  \bibinfo{person}{Ke Xu}, {and} \bibinfo{person}{Furu Wei}.}
  \bibinfo{year}{2020}\natexlab{}.
\newblock \showarticletitle{{BERT} Loses Patience: Fast and Robust Inference
  with Early Exit}. In \bibinfo{booktitle}{\emph{Advances in Neural Information
  Processing Systems}}, Vol.~\bibinfo{volume}{33}. \bibinfo{publisher}{Curran
  Associates, Inc.}, \bibinfo{address}{Red Hook, NY, USA},
  \bibinfo{pages}{18330--18341}.
\newblock


\bibitem[Zhu et~al\mbox{.}(2017)]%
        {zhu2017ocpc}
\bibfield{author}{\bibinfo{person}{Han Zhu}, \bibinfo{person}{Junqi Jin},
  \bibinfo{person}{Chang Tan}, \bibinfo{person}{Fei Pan},
  \bibinfo{person}{Yifan Zeng}, \bibinfo{person}{Han Li}, {and}
  \bibinfo{person}{Kun Gai}.} \bibinfo{year}{2017}\natexlab{}.
\newblock \showarticletitle{Optimized Cost per Click in {Taobao} Display
  Advertising}. In \bibinfo{booktitle}{\emph{Proceedings of the 23rd ACM SIGKDD
  International Conference on Knowledge Discovery and Data Mining}}.
  \bibinfo{publisher}{Association for Computing Machinery},
  \bibinfo{address}{New York, NY, USA}, \bibinfo{pages}{2191--2200}.
\newblock
\href{https://doi.org/10.1145/3097983.3098134}{doi:\nolinkurl{10.1145/3097983.3098134}}


\end{thebibliography}
\flushcolsend

\appendix
\clearpage
\section{Additional Details}
\label{sec:appendix}
\subsection{Model and Training Configurations}
\label{subsec:appendix-model-training}
The backbone uses EST tokenization, layer design, and prediction
readout. Behavioral sequences
are truncated in their existing order as depth increases;
non-behavioral tokens are retained. The two prediction depths have
independent task heads. Table~\ref{tab:model-configuration} lists the model
and training configuration.

\begin{table}[H]
\centering
\caption{Model and training configuration.}
\label{tab:model-configuration}
\small
\begin{tabular}{@{}p{0.44\columnwidth}@{\hspace{8pt}}p{\dimexpr0.56\columnwidth-8pt\relax}@{}}
\toprule
Field & Configuration \\
\midrule
Prefix / suffix depth $N$ / $M$ & 1 / 3 \\
Token hidden width $d$ & 128 (uniform across prefix and suffix) \\
Attention heads & 2 \\
FFN width & 256 ($128\rightarrow256\rightarrow128$) \\
Non-behavioral token groups & 32 \\
Task heads & Independent at the two depths \\
pGMV composition & pCVR $\times$ pAMT \\
LTR objective & COPR, Eq.~\eqref{eq:ltr-loss}; chunk size 10 \\
Value objectives & MSE on pCVR logits and $\log(1+\mathrm{pAMT})$ \\
Loss weighting & Fixed weights \\
Data split & Fine-ranking logs: 14 training days, 1 evaluation day \\
Optimizer / learning rate & Adam / 0.0001 \\
Training epochs / batch size & 1 / 20,000 per GPU \\
Training hardware & 128 H20 GPUs (96~GB each) \\
\bottomrule
\end{tabular}
\end{table}

In our implementation, pGMV is the product of pCVR, the purchase
probability given a click, and pAMT, the transaction amount given a
purchase. The value loss in Eq.~\eqref{eq:value-loss} applies MSE to
logit-transformed pCVR and $\log(1+\mathrm{pAMT})$, using the
corresponding transformed fine-ranking predictions as targets. The LTR
and value objectives share the same fine-ranking training examples.

The LTR objective follows COPR~\cite{zhao2023copr}. The LTR head outputs
$\hat{s}_i\in(0,1)$, and a relaxation factor $\alpha_i>0$, produced per
candidate by a small network, gives $\mathrm{pLTR}_i=\alpha_i\hat{s}_i$,
the signal used in Eq.~\eqref{eq:ecpm-ranking-score}. For each request,
the fine-ranking ECPM list is partitioned into fixed-size chunks and one
candidate is sampled from each chunk, forming $\mathcal{S}$; $\mathcal{P}$
in Eq.~\eqref{eq:ltr-loss} contains all pairs $(i,j)$ of sampled
candidates in which $i$ comes from a higher-ranked chunk than $j$. The
regularizer is $R=\sum_{i\in\mathcal{S}}r(\alpha_i)$, with
$r(\alpha)=\alpha-1$ for $\alpha>1$ and $r(\alpha)=1/\alpha-1$ otherwise.

\subsection{Serving and Cache Management Details}
\label{subsec:appendix-serving-cache}
The deployment uses two execution graphs, keeps the layer-$N$ states on
the GPU that produced them, and routes the retained candidates back to
that GPU for the deep round. Bidding parameters are read from a local
index in the first request and cached for the second. The bidding service
runs on GPUs in the same data center as prediction. To shorten the
online critical path, the prediction service exchanges pGMV predictions
and bids directly with the bidding service over a low-latency connection
and returns ECPM scores to the business engine. The engine aggregates
scores across packages and filters candidates rather than relaying
bidding traffic.
Table~\ref{tab:serving-configuration} summarizes the serving
configuration and cache-management settings.

\begin{table}[H]
\centering
\caption{Deployment and serving-benchmark configuration.}
\label{tab:serving-configuration}
\small
\begin{tabular}{@{}p{0.44\columnwidth}@{\hspace{8pt}}p{\dimexpr0.56\columnwidth-8pt\relax}@{}}
\toprule
Field & Configuration \\
\midrule
Candidate funnel & $16{,}000\to2{,}000\to400$ \\
Production packages & Up to 16 per request; up to 1,024 candidates each \\
Benchmark batch size & Production average batch-size distribution (up to 1,024 candidates) \\
Benchmark GPU & Pro5000, 72~GB, 1,345~GB/s \\
Compute precision & Mixed, identical in both graphs \\
Request deadline / cache TTL & 50 / 100~ms \\
Cache capacity per GPU & 15~GB (hard cap) \\
Fallback unit & Package \\
Mixed-depth scores & Direct ECPM merge, no calibration \\
\bottomrule
\end{tabular}
\end{table}

The cache holds the layer-$N$ states of every candidate in a package,
not only of the candidates later retained, because the states are
written before the engine filters candidates. Candidate-independent
user states are shared at package granularity, whereas candidate-dependent
states scale with the number of candidates. If candidate $i$ occupies
$n_i^{(N)}$ cached token slots of width $d$ at layer $N$ and each element uses
$b_{\mathrm{elem}}$ bytes, the candidate-dependent cached payload of request $q$ is
\begin{equation}
B_q = b_{\mathrm{elem}}\,d\sum_{i\in\mathcal{C}_q^{(0)}}n_i^{(N)}.
\label{eq:cache-payload}
\end{equation}
In our measured configuration, the cached states occupy about 24~KB per
candidate, or about 384~MB for a full request of 16,000 candidates. At the
maximum sustainable benchmark load of 1,950 packages/s per GPU and a 100~ms TTL,
taking the maximum package size of 1,024 candidates as an upper bound yields
the 4.8~GB resident-cache estimate in Section~\ref{subsec:cache-capacity-fallback}.

\end{document}